\documentclass[11pt,a4paper]{article}
\usepackage{amsmath,amssymb,amsthm,amsfonts}
\usepackage{geometry,graphicx}
\usepackage[dvipsnames]{xcolor}
\usepackage{hyperref}
\hypersetup{colorlinks=true,linkcolor=blue,citecolor=blue,urlcolor=blue}
\usepackage[numbers,sort&compress]{natbib}
\usepackage{bm}
\usepackage{authblk}

\usepackage{enumitem}

\newcommand{\bx}{\bm{x}}
\newcommand{\by}{\bm{y}}
\newcommand{\bz}{\bm{z}}
\newcommand{\bs}{\bm{s}}
\newcommand{\bv}{\bm{v}}

\newcommand{\bn}{\bm{n}}
\newcommand{\e}{\mathrm{e}}
\newcommand{\LR}{\mathrm{LR}}
\newcommand{\SR}{\mathrm{SR}}
\newcommand{\AT}{\mathrm{AT}}
\newcommand{\tsigma}{\tilde\sigma}
\newcommand{\tkappa}{\tilde\kappa}
\newcommand{\dd}{\mathrm{d}}
\newcommand{\Tr}{\operatorname{Tr}}
\newcommand{\cB}{\mathcal B}
\newcommand{\cP}{\mathcal P}
\newcommand{\post}{\mathrm{post}}
\newcommand{\Done}{\Delta_1}

\newcommand{\xhat}{\hat{\bm x}}
\newcommand{\Chat}{\hat{\mathbf C}}
\newcommand{\Egap}{E_{\rm gap}}
\newcommand{\lL}{\ell_\mathrm{L}}
\newcommand{\lD}{\ell_\mathrm{D}}
\newcommand{\DL}{D_\mathrm{L}}
\newcommand{\Gam}{\Gamma}

\title{Generative diffusion in long-range correlated landscapes:\\
Speciation cascade and the foam structure of the score}
\author[1]{J\'er\^ome Garnier-Brun}
\author[2]{Giulio Biroli}
\author[1,4]{Marc M\'ezard}
\author[3,4]{Jean-Philippe Bouchaud}
\affil[1]{Department of Computing Sciences, Bocconi University, Milan, Italy}
\affil[2]{Laboratoire de Physique de l'\'Ecole Normale Sup\'erieure, ENS, Paris, France}
\affil[3]{Capital Fund Management, Paris, France}
\affil[4]{Acad\'emie des Sciences, Paris, France}
\date{\today}

\begin{document}

\maketitle

\begin{abstract}
While state of the art for image and video generation, the fundamental reasons for the effectiveness of generative diffusion remain unclear. In particular, the observed advantage of following an entire reverse \textit{process} with respect to single-shot or autoregressive methods, and for what type of data it may hold, is not well understood. In this work, we aim to better understand the role of the reverse process by considering a complex yet analytically tractable data model, the Gibbs measure of a high-dimensional Gaussian random potential with long-range correlations. Focusing on the exact score, we show that the ``speciation'' threshold, in which a trajectory commits to one class of the data, becomes a speciation window, in which the reverse process progressively resolves the Parisi tree of pure states of the full replica-symmetry breaking measure, committing to ever finer branches down to a single pure state. Importantly, these commitments take place at noise levels where the noisy sample distinguishes the pure states only through an energy of order one, beyond the reach of standard replica saddle points. Constructing a Larkin argument suited to such $O(1)$ energy gaps, we describe the score as a \textit{foam}, with slab-like cells separated by shock-like boundaries. In the variance-exploding process we consider, this window is followed by a shrinkage of the samples towards the data support, and finally by reconstruction within the selected pure state. The denoising error of the exact score follows these stages, dropping during the speciation window with a profile set by the Parisi function alone, and then within the selected state. We believe that the generative foam we describe, and in particular the boundaries between cells, could play an important role in determining where the errors of trained diffusion models redistribute the mass of the generated samples when the data are long-range correlated.
\end{abstract}

\tableofcontents

\section{Introduction}
\label{sec:intro}

Diffusion-based models~\cite{sohl2015deep,ho2020denoising,song2020score} are state of the art in the artificial generation of images, videos and sounds. They operate by learning to reverse a noising process that maps a complex data distribution to an easy-to-sample counterpart, usually a Gaussian. Generation then occurs by following an entire stochastic \textit{trajectory} through the high-dimensional data space starting from pure noise and guided by a learned force field referred to as the score. Taking its roots in stochastic thermodynamics \cite{sohl2015deep}, this approach is particularly well suited to statistical physics analyses, which have notably been used to identify the different phases occurring in the reverse process \cite{biroli2023generative,raya2023spontaneous,biroli2024dynamical,achilli2026theory} and the transition of these models from generation to memorisation of training data \cite{achilli2024losing,ventura2024manifolds,achilli2025memorization,bonnaire2025diffusion}.

Understanding the role of data structure in generation is certainly a key open problem. In fact, one possible reason why diffusion models (DM) outperform earlier one-shot generative models, such as normalising flows \cite{rezende2015variational}, is that the time-dependent DM generation, in many practical applications, proceeds effectively scale by scale \cite{choi2022perception,guth2022wavelet}.  This makes them particularly well suited to multi-scale data such as natural images, videos and audio. Data are also hierarchically organised into abstract features and sub-features, which appears key to beating the curse of dimensionality~\cite{mezard2024spin,cagnetta2024deep}, and how this hierarchy emerges during generation is an active research topic~\cite{sclocchi2025phase,sclocchiprobing,favero2025compositional,garnier2026biased}. In these works, however, the hierarchy is a latent tree generating the elements of a single sample. Diffusion naturally raises the question of the \emph{spatial} structure \textit{between} samples, i.e. how they are organised relative to one another in data space, and how this organisation is recovered along the reverse trajectory \cite{klinger2026weighting}. This question then takes on a distinctly geometrical flavour, since structure in the data must show up as structure of a scalar landscape in the very space where sampling takes place, which is the core object of our paper.

As diffusion consists of progressively adding noise in ``time'' (written $t = 1/(2\kappa)$ in the variance-exploding process we consider below), this spatial recovery becomes a temporal one. In the simplest tractable model of structured data, a mixture of two Gaussians, \citet*{biroli2024dynamical} (BBBM) notably identified successive regimes of the reverse process. By reading the two modes as data categories, they defined and located a \emph{speciation time} at which a trajectory commits to one class. This criterion was since generalised by \citet{achilli2026theory} to arbitrary classes through Bayes classification, the speciation time being set by a free-entropy difference between classes, with successive speciation times when several classes are present. In a complementary, non-asymptotic approach, \citet{li2024critical} characterised {critical windows} of the reverse process, i.e.\ narrow time intervals during which a given feature of the output is decided, and interpreted diffusion as a hierarchical sampler that commits to features at a discrete sequence of times. In these models, however, structure is minimal to enable analytical descriptions, with a few classes or a finite hierarchy of mixtures specified by hand, so that speciation reduces to a finite sequence of commitments. 

Real data are instead correlated at all scales, likely organising into clusters within clusters. What becomes of speciation when it is no longer a few precise commitments but an entire cascade, and on which timescales the full hierarchy is recovered, remains open.
To address these questions, we focus on high-dimensional probability distributions borrowed from disordered systems theory, and which do show such full hierarchy. We study $N$-dimensional data drawn from a Gibbs measure $P_0(\bx)\propto\e^{-V(\bx)}$, where $V$ is a Gaussian random potential of the M\'ezard--Parisi type~\cite{mezard1990interfaces,mezard1991replica,mezard1992manifolds}, with $N$ large. More precisely, we consider the long-range correlated case $\overline{V(\bx)V(\by)}=B(0)-\sigma^2|\bx-\by|^{2H}$, $0<H<1$
, for which the Gibbs measure displays, at strong enough disorder, full replica-symmetry breaking (fRSB), i.e. an ultrametric tree of pure states embedded in space \cite{MPSTV84}. Such a setting is ``rough but not featureless'': rich enough to organise into a hierarchy of clusters, yet simple enough for closed-form results~\cite{mezard1990interfaces,mezard1991replica,engle1993replicas, le2008cusps}. In this work, we focus on the \emph{exact} score, in the vein of~\cite{biroli2024dynamical}; finite training sets on the same data landscape will be the object of a forthcoming work. Note that we are also not concerned here with the algorithmic implementation of such a denoiser, which is expected to be hard, at least in some regimes~\cite{elalaoui2022sampling,montanari2023sampling,huang2024sampling,ghio2024sampling,straziota2025generative}.

The Mézard–Parisi model lets us import decades of knowledge on disordered systems into the analysis of generative diffusion. As we shall see, however, the scales relevant to generation lie beyond the reach of the standard approach of a replica theory leading to an extensive saddle point. Indeed, in $N\gg 1$ dimensions, the multi-modal structure emerges at diffusion times where the signal-to-noise ratio is only of order $1/N$ per coordinate. As in the two-Gaussian analysis of BBBM, the sampling trajectory then commits to a branch of the hierarchy through the projection of the noisy point onto the few directions that separate competing modes, an $O(1)$ quantity, while the $N$ remaining directions only carry noise. The extensive saddle point, dominated by these $N$ directions, is blind to it. Note that this is not specific to our data model: whenever $P_0$ is dominated by a few macroscopically distinct states, the choice among them is made at these same times, where the noisy point exerts a total field of order one on the data. This is the diffusion counterpart of the vanishing symmetry-breaking field that selects a phase in the thermodynamic limit, and the limits of large $N$ and long diffusion times do not commute.

In order to describe the generative process, we therefore combine different techniques, all of which build on our knowledge of the Parisi tree of pure states and the statistics of their weights. We start by computing the covariance matrix of the data, allowing us to identify the scaling of the average onset of speciation following the spectral criterion introduced by BBBM (Sec.~\ref{sec:speciation}). Placing ourselves at this relevant timescale, we  propose an adapted Larkin balance argument~\cite{larkin1970effect,larkin1979pinning}, which recovers the speciation scaling and opens the description of a generative cascade that resolves the spatial structure level by level down the Parisi tree (Sec.~\ref{sec:cascade}). At intermediate times, this progressive refinement leads to a ``foam'' of cells separated by sharp walls---the shocks of an associated Burgers problem, that leads to the well known ``cusp'' singularity in the score correlator \cite{bouchaud1995scaling, balents1996large, le2008cusps} (Sec.~\ref{sec:shocks})---until the finest level is reached. 
We then translate this picture in terms of the Bayes-optimal denoising error, which follows the cascade down to the width of the selected state, before a planted replica computation shows that the recovery within that state at short times is a simple narrowing of the posterior with no further transition (Sec.~\ref{sec:attribution}). 
Finally, we argue that the walls of the foam are precisely where errors of an imperfect reverse dynamics feed excess probability (Sec.~\ref{sec:source}); as we will discuss in a forthcoming work, we conjecture that this abundance of well-organised shock surfaces is a mechanism that helps learned scores to generalise.

\subsection*{Our results in a nutshell: a multiscale speciation cascade}
\addcontentsline{toc}{section}{Our results in a nutshell: a multiscale speciation cascade}
\label{sec:summary}

\begin{figure}[t]
    \centering
    \includegraphics[width=\linewidth]{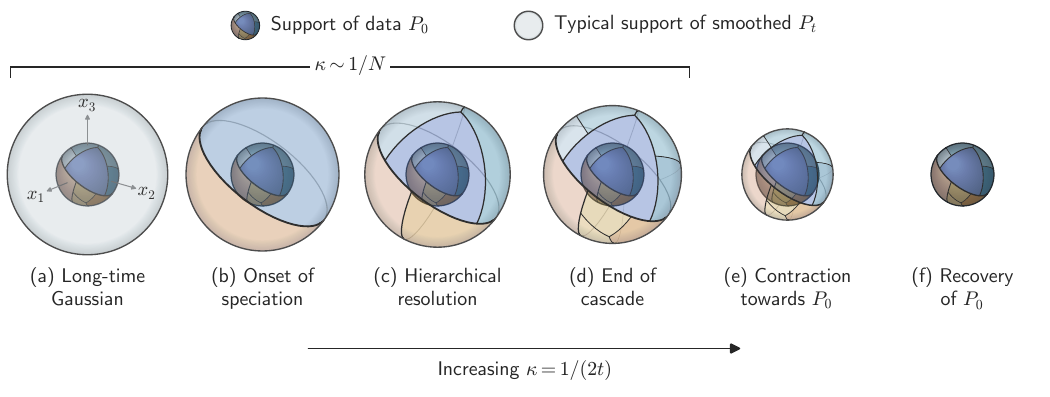}
    \caption{Schematic of hierarchical state resolution along the reverse diffusion scheme. The opaque inner sphere is the support of the data distribution $P_0$, of radius $L\sqrt{N}$, shown with its fixed partition into pure states; the translucent outer sphere marks the typical support of the smoothed distribution $P_t$, i.e. the shell of radius $\sqrt{N(L^2+\kappa^{-1})}$ on which it concentrates. (a) At the start of the reverse process $P_t$ is effectively Gaussian and the shell is homogeneous. (b) At the onset of speciation a single boundary divides it into two broad sectors, separated by a first shock. (c,d) Sectors are subdivided hierarchically until the shell carries the same cells as the data sphere, marking the end of the speciation cascade. (e) The shell then contracts onto the data sphere without further transition, keeping their global topology while modifying their local shapes, and (f) coincides with the data sphere as $t\to0$, recovering $P_0$. Colours label states and the branches that contain them, not computed probability densities. The figure is a schematic three-dimensional {cartoon} of an $N$-dimensional problem, which does not account for the very strong anisotropy of the cells in high dimension (see Sec.~\ref{sec:foam}).}
    \label{fig:two-sphere-sketch}
\end{figure}

Our model offers a generalisation of the speciation scenario of BBBM \cite{biroli2024dynamical} in the case of hierarchically structured data with long-range correlations. In such a case, speciation is not a single decision but a cascade of branchings down the Parisi tree of states, spread over a window of time scales all of order $N$ (or $\kappa = O(1/N)$) but separated by orders of magnitude when disorder is strong. 
We follow speciation through the evolution of the \emph{effective potential} $W$, which is defined as minus the logarithm of the probability distribution at time $t$ and whose derivative is minus the score function---the force which drives the generative dynamics. In the variance-exploding diffusion process $W$ reads
\begin{equation}
  {W(\by,\kappa)} =  - \log \int \dd\bx\,\e^{-V(\bx)-\frac\kappa2|\by-\bx|^2} - \frac{N}{2} \log \kappa, \qquad \kappa = \frac{1}{2t},
\end{equation}
which is the central object that we characterise geometrically in the following as a high dimensional, strongly anisotropic \emph{foam}. 
Our results are summarised in Fig.~\ref{fig:two-sphere-sketch}, see also Fig.~\ref{fig:eye-candy} below. As reverse diffusion operates, the foam structure of the effective potential subdivides and drives the speciation cascade from long to short diffusion time as 
\begin{itemize}
  \item \emph{Small $\kappa\to0$}. The smoothed potential $W(\by)$ has a single dominant basin;
  \item \emph{Speciation cascade}, $\tkappa_s/N\lesssim\kappa\lesssim\tkappa_c/N$ with $\tkappa_s \propto \tsigma$ and $\tkappa_c \propto\tsigma^{2/H}$, $0 < H < 1$, where $\tsigma = O(1)$ is the disorder strength appropriately rescaled with $N$ (Sec.~\ref{sec:potential}), such that the range of this regime diverges when $\tsigma \to \infty$. $W(\by)$ is a foam of slab-like cells of characteristic width $\lL \propto \kappa^{-1/(2-H)}$. The error in reconstructing a clean sample from its noised version decreases from its pure-noise value towards the width $\Done = O(1)$ (defined in Eq.~\eqref{eq:q1}) of a pure state that is assigned to a given trajectory;
  \item \emph{Selected state}, $\kappa\gg\tkappa_c/N$ but $\kappa\Done\ll1$, a window of width $\sim N$ in $\kappa$. The posterior distribution (or \emph{internal measure}) $ p(\bx \mid \by,\kappa)\propto \exp\left(-V(\bx) -\frac\kappa2|\by-\bx|^2\right)$ for typical $\by$ is a single pure state and the error stays close to $\Done$, so that no further transition happens; 
  \item \emph{Within the state}, $\kappa\Done\gtrsim1$. The posterior distribution can be shown to be replica symmetric and stable as it is confined within a state, $W(\by)$ approaches $V(\by)$, and the error decreases from $\Done$ to $1/\kappa$.
\end{itemize}
Note that the sharpness of the foam represented in Fig.~\ref{fig:eye-candy}, defined as the ratio $\Gam$ between the size of the cells and the thickness of the walls, is found to be given by
\begin{equation}
\Gam(\kappa):=\kappa\lL^2 \propto \kappa^{-H/(2-H)} .
\end{equation}
As a result, the cells resolved earlier in the cascade process are also the sharpest in that sense. Importantly, these walls are where excess probability is created when the reverse diffusion process is generated with errors in the learnt score (Sec.~\ref{sec:source}).

\section{Setting}
\label{sec:setting}

\subsection{Generative diffusion}
\label{sec:diffusion}

As mentioned in the introduction, diffusion models rely on transporting difficult to sample data points to an easy to sample measure, and to learn the reverse of that process in order to readily generate new samples. Here, we focus on a variance-exploding (VE) schedule, which allows for the simplest computations, although we expect our results to hold qualitatively in a variance-preserving (VP) setting.

The forward process is free diffusion from a data point,
\begin{equation}
  \dot{\by}(t)=\sqrt2\,\bm\eta(t),\qquad \by(0)=\bx\sim P_0,
  \label{eq:forward}
\end{equation}
i.e. $\by=\bx+\sqrt{2t}\,\bz$, with $\bz\sim\mathcal N(\bm0,\mathbf I_N)$. 
At the distribution level, this straightforwardly leads to a convolution between the initial distribution and a Gaussian:
\begin{equation} \label{eq:convolution}
  P_t=P_0\star\mathcal N(0,2t\,\mathbf I_N),
\end{equation}  
which is a solution to the heat equation $\partial_tP_t=\nabla^2P_t$. 

Generation runs the reverse process, which starting from $\by(T) \sim \mathcal{N}(\bm{0},2T\mathbf{I}_N)$ for $t=T\gg1$ can be written as the Langevin equation
\begin{equation}
  \dot{\by}(t)=-2\,\bs(\by,t)+\sqrt2\,\bm\eta(t),
  \label{eq:reverse}
\end{equation}
with the score
\begin{equation}
\bs(\by,t)=\nabla\log P_t(\by).
\end{equation}
Indeed, in reverse time $\tau=T-t$, the density of $\by$ then obeys the Fokker--Planck equation
\begin{equation}
  \partial_\tau P_\tau =-\nabla\cdot\big[2\bs\,P_\tau\big]+\nabla^2P_\tau,
  \label{eq:reverse-FP}
\end{equation}
which is solved by $P_{T-\tau}$, since $\bs P=\nabla P$.

In the following, we will often work with the ``\emph{precision}'' of the noise that we introduced above
\begin{equation}
  \kappa:=\frac1{2t},
  \label{eq:kappa}
\end{equation}
so that $\kappa\to0$ is pure noise and $\kappa\to\infty$ recovers the data. Note that in the Ornstein--Uhlenbeck (VP) scheme of~\cite{biroli2024dynamical} (which corresponds to keeping $\by$ on the sphere), $\bx_t=\e^{-t}\bx_0+\sqrt{1-\e^{-2t}}\,\bz$, the same signal-to-noise ratio corresponds to $\kappa=1/(\e^{2t_{\rm VP}}-1)$. In the large $N$ limit, the regime of speciation identified by BBBM, $t_{\rm VP}=\tfrac12\log(N)$, is therefore $\kappa=O(1/N)$, suggesting the rescaling
\begin{equation}
  \tkappa = \kappa N,
\end{equation}
a regime which will be at the heart of Secs.~\ref{sec:speciation}--\ref{sec:attribution}.

\subsubsection{Two measures}

In this work, the data are drawn from a Gibbs measure at unit temperature, $P_0(\bx)=\e^{-V(\bx)}/\mathcal Z$. Formally, this is not a restriction, since any distribution can be written in this form with the potential $V=-\log P_0$ up to a constant. In this language, we propose to distinguish two important objects. On one hand, we have the \emph{external} measure describing the distribution of the noisy data
\begin{equation}
  P_t(\by)=\left(\frac{\kappa}{2\pi}\right)^{N/2}\;\frac{1}{\mathcal{Z}}\;\int \dd\bx\,\e^{-V(\bx)-\frac\kappa2|\by-\bx|^2}, \quad \mathcal{Z}=\int \dd\bx \; \e^{-V(\bx)}\ .
  \label{eq:external}
\end{equation}
We will describe it with the fixed-$\by$ potential $W$, defined through
\begin{equation}
  Z(\by,\kappa):= \e^{-W(\by,\kappa)}= \left(\frac{\kappa}{2\pi}\right)^{N/2}\; \int \dd\bx\,\e^{-V(\bx)-\frac\kappa2|\by-\bx|^2}.
\end{equation}
On the other hand, we define the \emph{internal} measure,
\begin{equation}
  p(\bx \mid \by,\kappa)\propto \e^{-V(\bx)-\frac\kappa2|\by-\bx|^2}\ . 
  \label{eq:internal}
\end{equation}
This is the original measure of $\bx$ modified by the addition of a quadratic pinning field of precision $\kappa$ around $\by$. To understand the generation process of typical samples, we will be particularly interested in characterising the internal measure when $\by$ is a typical sample drawn from $P_t(\by)$. We will denote its mean by $\xhat(\by,\kappa):=\langle\bx\rangle_{\by,\kappa}$ and its covariance by $\Chat(\by,\kappa):=\langle\bx\bx^\top\rangle_{\by,\kappa}-\xhat\xhat^\top$.\footnote{Here and throughout the paper, $\langle \dots \rangle_{\by,\kappa}$ refers to the expectation over $p(\bx \mid \by,\kappa)$.} This internal measure has a straightforward interpretation in the diffusion setting, as the conditional mean $\xhat(\by,\kappa)$ is the Bayes-optimal denoiser given a sample corrupted by the Gaussian process at precision $\kappa$. It is directly related to the score through the exact Tweedie identity \cite{vincent2011connection}
\begin{equation}
  \nabla W(\by,\kappa)=-\bs=\kappa\big(\by-\xhat(\by,\kappa)\big).
  \label{eq:tweedie}
\end{equation}
This identity will also be relevant to study the geometry of the effective potential $W$, through the equality on the Hessian
\begin{equation}
  \nabla\nabla W(\by,\kappa)=\kappa\,\mathbf I_N-\kappa^2\,\Chat(\by,\kappa).
  \label{eq:hessian}
\end{equation}

Through the Tweedie identity, computing the score therefore amounts to solving a denoising problem. As such, it will be natural to characterise the reverse process through the denoising error of the Bayes-optimal $\xhat$, that is the minimum mean-squared error (MMSE),
\begin{equation}
  \Delta(\kappa):=\frac1N\,\mathbb E\,\big|\bx_0-\xhat(\by,\kappa)\big|^2, \qquad \by = \bx_0 + \frac{1}{\sqrt{\kappa}} \bz
  \label{eq:mmse}
\end{equation}
with $\mathbb E$ over $\bx_0\sim P_0$, the Gaussian white noise $\bz$, and the disorder. Since $\by$ is generated from $\bx_0\sim P_0$, Bayes' rule implies that given $\by$, $\bx_0$ is itself a sample of $p(\bx\mid\by,\kappa)$. This is the Nishimori identity, which allows the ground truth to be exchanged with a replica and will be essential to compute the MMSE analytically.

\subsubsection{A sharp limit: score shocks}
\label{sec:smoothing}

In this work, we choose to fix the temperature of the Gibbs measure to unity for convenience. However, in order to intuitively understand the geometry of $W$, it will be useful to think about a ``hardened'', zero temperature limit of the problem. For $\beta>0$ define
\begin{equation}
Z_\beta(\by,\kappa)=\left(\frac{\beta\kappa}{2\pi}\right)^{N/2}\int\dd\bx\,\exp\left(-\beta\left[V(\bx)+\frac\kappa2|\by-\bx|^2\right]\right), \qquad W_\beta=-\frac1\beta\log Z_\beta
\end{equation}
At $\beta=1$, $W_1\equiv W$ is the smoothed potential of the diffusion. As $Z_\beta$ solves the heat equation with diffusion constant $1/\beta$, a Cole-Hopf transform leads $W_\beta$ to follow a deterministic KPZ equation (see e.g. \cite{bouchaud1995scaling}) with viscosity $1/\beta$,
\begin{equation}
  \partial_tW_\beta=\frac1\beta\nabla^2W_\beta-|\nabla W_\beta|^2,\qquad W_\beta(\by,0)=V(\by),
  \label{eq:KPZ}
\end{equation}
and thus $\bv = 2\nabla W_\beta$ is a Burgers velocity field \cite{frisch2002burgulence, bec2007burgers} (also called the Polchinski equation in the context of the renormalisation group, see e.g. \cite{bauerschmidt2024stochastic}). We can thus identify $\bs = -\frac12 \bv$, extending the mapping discussed in~\citet{sarkar2026score} to $N$ dimensions.

The inviscid limit $\beta \to \infty$ leads to the Hopf--Lax solution (also called the Moreau envelope of $V(\bx)$)
\begin{equation}
  W_\infty(\by,t)=\min_{\bx}\left[V(\bx)+\frac{|\by-\bx|^2}{4t}\right],
  \label{eq:hopf-lax}
\end{equation}
that is the zero-temperature M\'ezard--Parisi problem~\cite{mezard1990interfaces,balents1996large} and inviscid Burgers with random initial data~\cite{bouchaud1995scaling}. It has genuine shocks where the minimiser jumps, which are rounded at finite $\beta$. Note that since $\beta V$ has the statistics of $V$ at a stronger disorder (Sec.~\ref{sec:potential}), this sharp limit is also the one approached at strong disorder. We shall study this shock-like picture in Secs.~\ref{sec:foam} and \ref{sec:shocks}.

\subsection{The data: a long-range random potential and its Parisi tree}
\label{sec:potential}\label{sec:equilibrium}

We now detail the data model we are considering in this study. We take $P_0(\bx)=\e^{-V(\bx)}/\mathcal{Z}$ on the sphere $\Omega=S^{N-1}(L\sqrt N)$ (a hard-wall ball, cube or torus of side $L\sqrt N$ can be used interchangeably), with $V$ a centred Gaussian potential of covariance
\begin{equation}
  \overline{V(\bx_1)V(\bx_2)}=B(|\bx_1-\bx_2|)=B(0)-\sigma^2|\bx_1-\bx_2|^{2H},\qquad 0<H<1,
  \label{eq:cov-function}
\end{equation}
in the inertial range $\xi_\SR\ll|\bx_1-\bx_2|\ll\xi_\LR$, with $\xi_\SR = O(1)$ a short-range ``ultraviolet'' regulariser and $\xi_\LR$ the long-range cutoff giving the largest correlation length, that we will choose to be equal to the size of the ball. Overlines will henceforth indicate averages over the realisation of the potential, and the temperature is set to one.\footnote{Since $\beta V$ has covariance $\beta^2B$, a temperature $1/\beta$ would only amount to $\sigma\to\beta\sigma$, so that the disorder strength plays the role of an inverse temperature and setting the temperature to one loses no generality.} The Hurst exponent $H\in(0,1)$ controls the roughness: $H \to 0$ corresponds to log-correlated-like landscapes~\cite{fyodorov2008freezing,fyodorov2008statistical}; $H\to 1$ approaches smooth quadratic covariance. The case $H=1/2$ in dimension $N=1$ is a Brownian potential which has received a lot of attention \cite{mezard1992manifolds,engle1993replicas,broderix1995thermal,comtet1998exponential}.

Configurations on the sphere are typically $O(\sqrt N)$ apart. For the large-$N$ limit to be non-trivial, these distances must lie inside the inertial range, and the variance of the potential across them must be extensive, so that energy and entropy compete at the same order. We thus adopt the canonical large-$N$ scaling
\begin{equation}
  \sigma=\tsigma\,N^{(1-H)/2},
  \label{eq:canonical}
\end{equation}
with $\xi_\LR=L\sqrt N$ and $L,\tsigma=O(1)$, under which the large scale square excursion of the potential $\tsigma^2L^{2H}N$ is extensive in $N$. Note that all results shall depend on $\tsigma$ and $L$ only through $\tsigma L^H$, once lengths are measured in units of $L$. 

In the large-$N$ limit, $P_0$ can be studied by the replica method and is described by the overlaps $Q_{ab}=\bx_a\cdot\bx_b/N$ between configurations. For geometric statements we use instead squared distances per coordinate,
\begin{equation}
  D_{ab}^2=\frac{|\bx_a-\bx_b|^2}{N}=2(L^2-Q_{ab}),
  \label{eq:dictionary}
\end{equation}
together with the physical length $\ell=\sqrt{N}D$ and the width $\Delta=D^2/2$. In these variables the covariance reads $B(\sqrt{N}D)=N\cB(D^2)$, with $\cB(z)=B(0)/N-\tsigma^2z^{H}$. 

Let us summarise the most salient features of the Gibbs measure $P_0$, obtained with the replica method~\cite{mezard1990interfaces,mezard1991replica}; we refer unfamiliar readers to Appendix~\ref{app:replica}, where we rederive these results in our notation following~\cite{fyodorov2007classical}. The first important result is that the replica-symmetric solution $Q_{ab}=q$ for all $a\neq b$ is stable only for $\tsigma<\tsigma_\AT$, where
\begin{equation}
  \tsigma_\AT^2L^{2H}=\frac1{2^HH(1-H)\,[(1-H)/(2-H)]^H}.
  \label{eq:sigma-AT}
\end{equation}
Since we want a highly structured distribution, we place ourselves beyond this de~Almeida--Thouless line, $\tsigma>\tsigma_\AT$, where the measure undergoes full replica symmetry breaking (fRSB). We call this regime strong disorder, and will further typically assume $\tsigma\gg1$ for the asymptotic statements.

The Parisi function $u(q)$, i.e. the probability that two configurations drawn from $P_0$ have an overlap smaller than $q$, is then continuous between the two plateau values $q_0$ and $q_1$ of the overlap,
\begin{equation}
  u(q_0 < q < q_1)=\frac{2-H}{2\tsigma\sqrt{H(1-H)}}\,D^{-H}, \quad \text{with} \;\; {D^2=2(L^2-q)}.
  \label{eq:xq}
\end{equation}
The upper plateau sets the width of a pure state,
\begin{equation}
  \Delta_1=L^2-q_1=\left[\frac1{2^H\tsigma^2H(1-H)}\right]^{1/H},
  \label{eq:q1}
\end{equation}
which shrinks as $\tsigma^{-2/H}$ at strong disorder, while the lower plateau,
\begin{equation}
  q_0=\frac{L^2}{2-H},
  \label{eq:q0}
\end{equation}
i.e. $D_0=L \sqrt{2(1-H)/(2-H)}$, does not depend on the disorder strength. At the two ends, $u$ takes the values $u_0=u(q_0^+)\propto1/\tsigma$ and $u_1=u(q_1^-)=(2-H)/2$, the latter independent of $\tsigma$. It will be convenient to label the levels of the hierarchy by the Parisi level $u(q)$, and to invert the relation:
\begin{equation}
  D(u)=\left[\frac{2-H}{2\tsigma\sqrt{H(1-H)}}\right]^{1/H}u^{-1/H},\qquad u_0<u<u_1 .
  \label{eq:Du}
\end{equation}

Physically, $P_0$ decomposes into pure states $\alpha$, i.e. its ergodic components, with weights $w_\alpha\propto\e^{-F_\alpha}$, centres $\bm m^\alpha$ and thermal width $\Delta_1$ per coordinate. Two states drawn from $P_0$ are at a distance of at least $D(u)$ with probability $u$~\cite{parisi1983order}, so that the physical distance between them is distributed as
\begin{equation}
  \rho(\ell)\propto\ell^{-1-H},\qquad \ell_1<\ell<\ell_0,
  \label{eq:rho-ell}
\end{equation}
with atoms of weight $u_0$ at $\ell_0=\sqrt{N}D_0$ and $H/2$ at $\ell_1=\sqrt{N}D_1$, the latter corresponding to two configurations in the same state. The exponent is fixed by the roughness of $V$ alone. Since $\ell_0=O(\sqrt N)$ and $\ell_1\propto\sqrt N\,\tsigma^{-1/H}$, the whole hierarchy lies inside the inertial range as $N\to\infty$, and strong disorder makes it broad, $D_1/D_0\propto\tsigma^{-1/H} \to 0$.

The states are organised in an ultrametric tree, the Parisi tree of states (M\'ezard--Virasoro)~\cite{MPSTV84,mezard1985microstructure} (App.~\ref{app:A-weights}). Two states whose branches separate at level $q$ have overlap $q$. At each level, the free energies of the competing sub-clusters behave as the low-lying levels of Derrida's random energy model~\cite{derrida1981random}. They are independent, with a density that grows exponentially as $\e^{u(q)F}$~\cite{MPSTV84,mezard1985random,mezard1985microstructure}, so that only a few of them carry a significant weight. At the finest level, the participation ratio of the pure states is~\cite{MPSTV84,derrida1985sample}
\begin{equation}
  \overline{\sum_\alpha w_\alpha^2}=1-u_1=\frac H2 .
  \label{eq:participation}
\end{equation}
Although infinitely many states contribute as $N\to\infty$, a few of them carry most of the weight, and the measure is condensed.

\subsection{Pure state decomposition of the noised measure}
\label{sec:potential-decomp}

The decomposition of $P_0$ into pure states carries over to $P_t$. Writing $P_0=\sum_\alpha w_\alpha P_0^{(\alpha)}$, with $P_0^{(\alpha)}$ the measure restricted to the state $\alpha$ and $w_\alpha=\e^{-F_\alpha}/\mathcal Z$, the convolution~\eqref{eq:convolution} gives
\begin{equation} \label{eq:purestate}
  P_t(\by)=\sum_\alpha w_\alpha\,\big(P_0^{(\alpha)}\star\mathcal N(0,2t\,\mathbf I_N)\big)(\by)=\frac1{\mathcal Z}\sum_{\alpha}\e^{-F_\alpha(\by)},
\end{equation}
where $F_\alpha(\by)=F_\alpha-\log\big(P_0^{(\alpha)}\star\mathcal N(0,2t\,\mathbf I_N)\big)(\by)$ is the free energy of the state $\alpha$ pinned around $\by$. Since $\e^{-W}=\mathcal Z P_t$, the smoothed potential is exactly a soft minimum over the states,
\begin{equation}
  W(\by,\kappa)=-\log\sum_\alpha\e^{-F_\alpha(\by)} .
  \label{eq:W-softmin}
\end{equation}

Even though the initial measure $P_0$ is fragmented in pure states, by increasing $t$ the noised versions of the pure states probability distributions merge leading to a cascade of transitions and eventually to a unique pure state for $P_t$---coinciding with a Gaussian when $t\rightarrow \infty$. Understanding this cascade is the aim of the next section. We will do it focusing on the potential $W$ which is equal (up to a constant) to minus the log of $P_t$.

\section{The onset of speciation}
\label{sec:speciation}

\subsection{The speciation time}
\label{sec:speciation_def}

One of the key results of BBBM is the identification of the so-called \textit{speciation time} $t_s$. Strictly speaking, there are in fact two possible definitions of $t_s$.

The first, which motivates its name, is one of \textit{attribution}. Assuming a distribution is multimodal and each mode can be identified with a class, it is natural to define $t_s$ as the time at which a given trajectory becomes committed to one of the classes. \citet{achilli2026theory} makes this precise through Bayes classification of the noised configuration, locating $t_s$ where the free-entropy difference between two classes becomes comparable to its fluctuations.

The second is instead based on \textit{geometry}. As the long-time distribution is a unimodal Gaussian, one can ask when non-trivial modes start to emerge in the potential $W$. This question itself has two answers: (i) the time at which the long-time basin of $W$ loses convexity, a spinodal; and (ii) the time at which new stationary points of $W$ first appear away from it. The two need not coincide, and in between may lie a binodal region in which the long-time basin is still convex while other basins already exist, as in a first-order transition. We come back to this distinction in Sec.~\ref{sec:cascade}.

\subsection{A spectral criterion}

In the mixture of two Gaussians of equal weights studied in BBBM, all of these coincide: the loss of convexity is a pitchfork bifurcation that creates the two new basins, and trajectories commit at the same time. The speciation time can then be located by the simplest of these criteria, the loss of convexity of the long-time basin, which takes the form of a spectral criterion.

While initially proposed in the variance-preserving context, the argument only uses loss of convexity and therefore transposes directly to our variance-exploding setting. Let $\by^\star$ be the unique minimum of $W$ at long times. By Eq.~\eqref{eq:tweedie}, $\by^\star=\xhat(\by^\star,\kappa)$, and by Eq.~\eqref{eq:hessian} it destabilises when
\begin{equation}
  \kappa_s\,\hat c_{\max}(\by^\star,\kappa_s)=1,
  \label{eq:spectral}
\end{equation}
where $\hat c_{\max}$ is the largest eigenvalue of the posterior covariance $\hat{\mathbf{C}}$. Now, assume that this occurs at small $\kappa$ (which we can check \textit{a posteriori}). It then follows that $\Chat$ is well approximated by the data covariance $\mathbf C$ of $P_0$, giving the spectral criterion $t_s\simeq c_{\max}(\mathbf C)/2$. Therefore the long time behaviour of the reverse process can be predicted by the bare covariance matrix of the data.

\subsection{Data covariance of the fRSB measure}
\label{sec:covariance}

We now set out to explicitly compute the speciation time for our data model. We start with a pure state decomposition,
\begin{equation}
  \mathbf C=\underbrace{\sum_\alpha w_\alpha\mathbf C^\alpha}_{\mathbf C_{\rm intra}}+\underbrace{\sum_\alpha w_\alpha(\bm m^\alpha-\bar{\bm m})(\bm m^\alpha-\bar{\bm m})^\top}_{\mathbf C_{\rm inter}},
  \label{eq:C-decomp}
\end{equation}
which we have itself decomposed in a intrastate and an interstate contribution. Here $\bm m^\alpha=\langle\bx\rangle_\alpha$ is the mean within a pure state, $\bar{\bm m}=\sum_\alpha w_\alpha\bm m^\alpha$, and $C^\alpha_{ij}=\langle x_ix_j\rangle_\alpha-\langle x_i\rangle_\alpha\langle x_j\rangle_\alpha$ describes intrastate fluctuations, while $\mathbf C_{\rm inter}$ represents interstate contributions.

We start by computing the interstate covariance. For any two states, the study of the microstructure of ultrametricity of \citet{mezard1985microstructure}, adapted to a Gaussian measure, gives
\begin{equation}
  \overline{m_i^\alpha m_j^\gamma\mid Q_{\alpha\gamma}=q}=\delta_{ij}\,q,
  \label{eq:C-state-means}
\end{equation}
with $q=q_1$ for $\gamma=\alpha$. Therefore, when averaging over the disorder,
\begin{equation}
  \frac1N\overline{\Tr\, \mathbf C_{\rm inter}}=q_1-\int_0^1\dd u\,q(u)=q_1-\bar q,
  \label{eq:C-trace}
\end{equation}
with $\bar q=\int\dd q\,P(q)\,q$ the mean overlap, and $P(q)=\dd u/\dd q$ the distribution of overlaps, which includes peaks of weight $u_0$ at $q_0$ and $1-u_1$ at $q_1$. Rotational invariance then implies
\begin{equation}
  \overline{(\mathbf C_{\rm inter})_{ij}}=\Delta_m\,\delta_{ij},
  \label{eq:C-Delta-m-def}
\end{equation}
with $\Delta_m:=q_1-\bar q$. Integrating by parts and using the fRSB solution of Sec.~\ref{sec:potential},
\begin{equation}
  \Delta_m=\int_{q_0}^{q_1}u(q)\,\dd q=\frac{u_0D^2_0-u_1D^2_1}{2-H}=\frac{D_0^{2-H}-D_1^{2-H}}{2\tsigma\sqrt{H(1-H)}}\;\propto\;\frac1{\tsigma}.
  \label{eq:Delta-m}
\end{equation}

The above diagonal expression is only true on average, as it suppresses the random orientations of possible large eigenvectors. Let $\mathbf M$ be the $N\times\mathcal N_s$ matrix whose columns are the vectors $\bm m^\alpha$, let $\bm w$ be the column vector of state weights, and set $\mathbf A_{\bm w}=\operatorname{diag}(\bm w)-\bm w\bm w^\top$. Then $\mathbf C_{\rm inter}=\mathbf M\mathbf A_{\bm w}\mathbf M^\top$, $\frac1N\mathbf M^\top\mathbf M=\mathbf Q$, the latter being the full matrix of overlaps $Q_{\alpha\gamma}=\bm m^\alpha\cdot\bm m^\gamma/N$ between state centres. Hence the nonzero spectra obey
\begin{equation}
  \operatorname{Spec}_{\neq0}(\mathbf C_{\rm inter})=N\operatorname{Spec}_{\neq0}(\mathbf G),
  \label{eq:C-Gram}
\end{equation}
with $\mathbf G=\mathbf A_{\bm w}^{1/2}\mathbf Q\,\mathbf A_{\bm w}^{1/2}$. This reduces the required spectral problem to the overlap matrix and the statistics of the weights.

In the simplest case of two states, $\mathbf G$ has rank one and Eq.~\eqref{eq:C-Gram} gives the unique nonzero eigenvalue $2Nw(1-w)(q_1-Q_{\alpha\gamma})=Nw(1-w)D^2_{\alpha\gamma}$. In the fRSB case, the effective number of states contributing to the measure is given by the participation ratio, which as mentioned above is $O(1)$ [Eq.~\eqref{eq:participation}]. This leads us to assume that the effective rank
\begin{equation}
  r_{\rm eff}(\mathbf G)=\frac{\Tr\mathbf G}{\lambda_{\max}(\mathbf G)}=\frac{\Tr\mathbf C_{\rm inter}}{c_{\max}(\mathbf C_{\rm inter})}
\end{equation}
remains $O(1)$ as $N\to\infty$, and carries no power-law dependence on $\tsigma$. 
While the participation ratio is known to fluctuate from sample to sample, the scaling of the ratio for a typical sample is the same as the one of the disordered average (as we check below). Using  Eq.~\eqref{eq:C-Gram} we thus get
\begin{equation}
  c_{\max}(\mathbf C_{\rm inter})=N\lambda_{\max}(\mathbf G)\sim Ng_H\Delta_m\propto\frac N\tsigma,
  \label{eq:C-cmax-inter}
\end{equation}
where $g_H=O(1)$ incorporates the effective rank and the sample-to-sample fluctuations of $\Tr\mathbf G$. 

We finally consider the connected fluctuations inside the pure states, $\mathbf C_{\rm intra}=\sum_\alpha w_\alpha\mathbf C^\alpha$. For a typical state, we write $\mathbf K=(\mathbf C^\alpha)^{-1}$ the stiffness matrix. As the fRSB solution is marginally stable, we assume that this marginality appears in the physical intrastate susceptibility through a soft lower edge of the normalised stiffness spectrum,
\begin{equation}
  \rho_K(\lambda)=f\lambda^\theta+o(\lambda^\theta),\qquad \lambda\to0,\qquad 0<\theta<1.
  \label{eq:C-soft-edge}
\end{equation}
At finite $N$, the typical smallest stiffness is estimated by requiring an order-one number of eigenvalues below it, $N\int_0^{\lambda_{\min}}\dd\lambda\,\rho_K(\lambda)\simeq1$, which gives
\begin{equation}
  \lambda_{\min}\simeq\left(\frac{1+\theta}{Nf}\right)^{1/(1+\theta)},
  \label{eq:C-lambda-min}
\end{equation}
with $c_{\max}(\mathbf C^\alpha)=\lambda_{\min}^{-1}$. The density of soft modes $f$ is fixed by the width of the states, $N^{-1}\Tr\mathbf C^\alpha=\Done$, under a one-scale assumption on the stiffness spectrum (App.~\ref{app:C-intra}), so that
\begin{equation}
  c_{\max}(\mathbf C^\alpha)\ \propto\ N^{1/(1+\theta)}\tsigma^{-2/H}.
  \label{eq:C-cmax-intra}
\end{equation}
The same scaling bounds the weighted intrastate contribution, since $c_{\max}(\mathbf C_{\rm intra})\leq\sum_\alpha w_\alpha c_{\max}(\mathbf C^\alpha)$. For every $\theta>0$, this contribution is subextensive in $N$ and is therefore subdominant to the $O(N)$ interstate eigenvalue.

Combining the two contributions, the spectral criterion~\eqref{eq:spectral} gives
\begin{equation}
  \tkappa_s\propto\tsigma \quad \longrightarrow \quad t_s\propto \frac{N}{\tsigma}.
  \label{eq:ts}
\end{equation}
Since $\kappa_s=O(1/N)$, using the $\kappa=0$ covariance is consistent at the level of scaling. 

Let us interpret the role of the disorder strength. Here, the dependence does not come from the width of the states, whose narrowing, $\Delta_1\propto\tsigma^{-2/H}$, only affects the subdominant intrastate term. Instead, it comes from the spread of the state centres, $\Delta_m=q_1-\bar q$, which is half the mean squared distance per coordinate between the centres of two states drawn from $P_0$. At strong disorder this spread decreases as $1/\tsigma$, because the Gibbs weight concentrates on few, closely related states. Two samples of the data then typically belong to the same deep branch, even though the hierarchy extends up to distances $D_0=O(1)$. Resolving this smaller spread requires less noise, and speciation occurs at larger $\kappa$, i.e. later in the reverse process.

As mentioned above, however, this estimate is annealed due to the disorder-averaged covariance, through the assumption that the trace of $\mathbf C_{\rm inter}$ in a typical sample scales as its average. Since the weights of the pure states fluctuate strongly from one sample to the next, the measure being condensed (Sec.~\ref{sec:potential}), it remains to check whether Eq.~\eqref{eq:ts} holds for a given sample. We address this question in the following section.

\section{Speciation cascade and generative foam}
\label{sec:cascade}

Section~\ref{sec:speciation} located the onset of speciation at $\tkappa_s\propto\tsigma$, i.e. at $\kappa=O(1/N)$, consistent with the work of BBBM and \citet{achilli2026theory}. The challenge with this scale in our disordered setting where the score cannot be written explicitly is that the part of the pinning that distinguishes two states $\alpha$ and $\gamma$ in $W$, $\kappa\,\by\cdot(\bm m^\alpha-\bm m^\gamma)$, is of order one, whereas the free energy, and with it the replicated action, is of order $N$. A standard replica computation, based on an extensive saddle point,  cannot see the competition between states and is therefore not suitable to study this long-time regime. 

\begin{figure}[t]
    \centering
    \includegraphics[width=\linewidth]{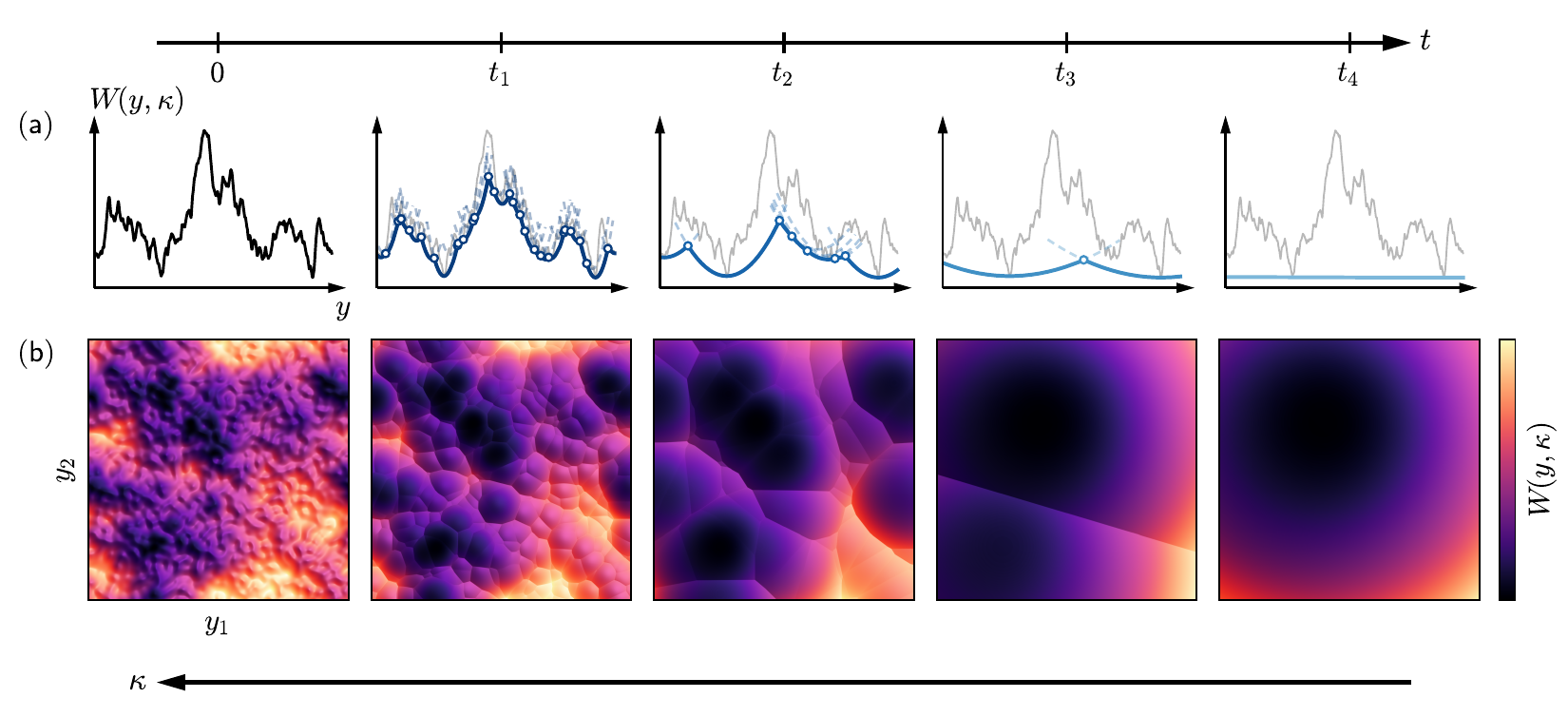}
    \caption{Smoothing of a rough landscape with $H = 3/4$ under the variance-exploding diffusion process in (a) $N= 1$ and (b) $N=2$ dimensions. The smoothed potential $W$ is shown in its zero-temperature (Hopf--Lax) form, Eq.~\eqref{eq:hopf-lax}, for easier visualisation. In (a), markers show shocks, where the minimiser $\bx^\star(\by)$ of the Hopf--Lax solution jumps between two competing basins and $W$ has a kink; dashed lines continue each branch a short distance beyond its shock. For (b), we show a hill-shaded $W$, with the colour scale normalised in each panel (dark: low $W$, i.e. high $P_t$). Shocks appear as creases separating the cells of a ``foam''. As $t$ grows the cells merge hierarchically, from a dense foam ($t_1$) to a few large basins ($t_2$) and a single remaining shock ($t_3$), until one basin fills the box ($t_4$). The times are chosen separately for the two rows to show comparable stages.}
    \label{fig:eye-candy}
\end{figure}

Two questions remain open. The spectral estimate of Sec.~\ref{sec:speciation} is annealed, and an independent, sample-wise derivation of the onset is needed. Moreover, unlike the two-component mixture~\cite{biroli2023generative,biroli2024dynamical}, the fRSB measure contains a hierarchy of competing pure states. It is thus interesting to follow how the reverse ``cascade'' process resolves it beyond the first split. Increasing $\kappa$, the reverse process eventually needs to resolve finer branches of the Parisi tree. At intermediate scales, it appears natural that $W(\by,\kappa)$ is locally dominated by different states in different regions of $\by$-space. We call the resulting collection of basins, separated by shock-like walls, the \emph{generative foam}, see Fig.~\ref{fig:eye-candy} for a low-dimensional illustration.

In this section, we address both questions with a Larkin argument, which balances the $O(1)$ pinning against the free-energy gaps between branches of the Parisi tree and holds sample by sample (Sec.~\ref{sec:larkin}). It gives an independent estimate of the onset, follows the resolution down the tree (Sec.~\ref{sec:cursor}), and fixes the geometry of the foam (Sec.~\ref{sec:foam}). 

\subsection{The Larkin balance}
\label{sec:larkin}
The Larkin argument was first introduced to study the collective pinning of vortex lattices~\cite{larkin1970effect,larkin1979pinning}. Our problem is its analogue on a $N$-dimensional sphere~\cite{mezard1992manifolds}, as one configuration $\bx$ in the random potential $V$ is confined around $\by$ by $\kappa|\by-\bx|^2/2$. Beyond the Larkin scale several metastable configurations compete, and the preferred one jumps as the centre $\by$ of the well moves, producing a shock in the ``zero temperature'' Hopf--Lax limit~\cite{bouchaud1995scaling}.

The Larkin balance can therefore be applied, but its usual geometric interpretation must be adapted to the diffusion setting in large dimensions. Indeed, at the speciation scale, $\kappa=\tkappa/N$, a typical noised configuration has $|\by|=O(N)$, whereas the competing states remain on the sphere $|\bx|=L\sqrt N$. The distance $|\by-\bx|$ is consequently dominated by the noise, 
such that we need to be careful in comparing the pinned free energies~\eqref{eq:purestate} of two competing pure states $\alpha$ and $\gamma$ of $P_0$, with centres $\bm m^\alpha$ and $\bm m^\gamma$ and free energies $F_\alpha$ and $F_\gamma$. 


In the following we will make a \emph{rigid-state} approximation
\begin{equation}
    F_\alpha({\by})=F_\alpha-\log \left\langle\e^{-\frac{\kappa}{2}|\by-\bx|^2}\right\rangle_\alpha\simeq F_\alpha+\frac{\kappa}{2}\left\langle|\by-\bx|^2\right\rangle_\alpha=F_\alpha+\frac\kappa2|\by-\bm m^\alpha|^2 +{\rm const},
  \label{eq:rigid-state}
\end{equation}
and likewise for $F_\gamma({\by})$, with $\langle\cdot\rangle_\alpha$ the average within the state $\alpha$. This is a good approximation because the $\bx$-dependent terms of the pinning field lead to non-extensive contributions of order one in the speciation regime. As such, they can be treated perturbatively, thus justifying the expansion above (see also \cite{biroli2023generative}), while the intra-state fluctuations are, to leading order, the same for all states at a given $\by$. Starting from this expression, we can compare the two contributions. Since the two states have the same norm $|\bm m^\alpha|^2=N q_1$ to leading order in $N$, their relative free energy is
\begin{equation}
  F_\alpha(\by)-F_\gamma(\by)=F_\alpha-F_\gamma-\kappa\,\by\cdot\bm d ,
  \label{eq:two-state}
\end{equation}
with $\bm d=\bm m^\alpha-\bm m^\gamma:= \ell \bm n$ and $|\bm n|=1$. All terms involving $|\by|^2$, as well as the equal radial contributions from the two states, cancel from this difference.

Let us decompose $\by$ as $\by=\phi \ell \bm n+\by_\perp$, with $\by_\perp\cdot\bm d=0$, such that the latter contribution also disappears from Eq.~\eqref{eq:two-state}. Here $\phi=O(1)$ is the position of $\by$ along the axis joining the two centres, with---say---$\phi=\frac12$ when $\by = \bm m^\alpha$ and  $\phi=-\frac12$ when $\by = \bm m^\gamma$. Equation~\eqref{eq:two-state} then gives the crossing condition $\kappa\phi\ell^2 = \Egap(\ell)$, where $\Egap(\ell)$ is the typical free-energy difference between the deepest state within distance $\ell$ and its closest \textit{competitor}. Since $\phi = O(1)$, this defines an
operative \emph{Larkin scale} such that
\begin{equation}
  \kappa\,\lL^2\sim\Egap(\lL).
  \label{eq:larkin-balance}
\end{equation}
The energy balance has the familiar Larkin form, but here it arises only after projecting the $O(N)$ noised configuration onto the $O(\sqrt N)$ direction separating two competing states.

The subtle point now is to estimate the relevant energy gap in high dimension. There, a shell at distance $\ell$ from a reference minimum can be divided into $\mathcal M\sim\e^{cN}$ macroscopically distinct cells, whose mutual separations are an $O(1)$ fraction of $\ell$. This exponential multiplicity is consistent with complexity calculations for high-dimensional random landscapes~\cite{bray2006statistics,fyodorov2007classical}. This count is much larger than in low dimensions but much smaller than the naive $\ell^N$ count that ignores correlations. Treating the macro-cells as effectively independent, extreme-value statistics of $\mathcal M$ Gaussian variables of variance $\sigma^2\ell^{2H}$ then gives
\begin{equation}
  \Egap(\ell)\sim\frac{\sigma\ell^H}{\sqrt N}=\tsigma\left(\frac\ell{\sqrt N}\right)^H .
  \label{eq:gap}
\end{equation}
Here, the $1/\sqrt N$ suppression relative to the naive $\sigma\ell^H$ is the crucial correction: in many dimensions, low-lying states are exponentially numerous, and their gaps are correspondingly compressed by an extreme-value factor $1/\sqrt{2\log\mathcal M}\sim1/\sqrt{cN}$.

Importantly, this scaling is consistent with the replica solution. First, at large scales, i.e. for the deeper states corresponding to $\ell \sim \sqrt{N}$, we find an energy gap of order one as expected in fRSB. Second, and more precisely, at temperature $T$, the analogue of Eq.~\eqref{eq:xq} reads~\cite{fyodorov2007classical}
\begin{equation}
  u_T(q)=\left.\frac T2\,\frac{2-H}{\tsigma\sqrt{H(1-H)}}\,D^{-H}\right|_{D^2=2(L^2-q)} .
\end{equation}
At a given level of the Parisi hierarchy, $u_T(q)$ sets the characteristic free-energy gap  (Sec.~\ref{sec:potential}),
\begin{equation}
  \Egap(D)\sim\frac{T}{u_T(D)}\sim\tsigma D^{H}.
  \label{eq:gap-x}
\end{equation}
The gap is therefore independent of the temperature, and using $D=\ell/\sqrt{N}$ recovers Eq.~\eqref{eq:gap}. The same gap can also be reproduced without replicas, in a generalised random energy model built in real space from a dyadic decomposition of $V$ (App.~\ref{app:gaps-grem}).

\begin{figure}[t]
    \centering
    \includegraphics[width=0.54\linewidth]{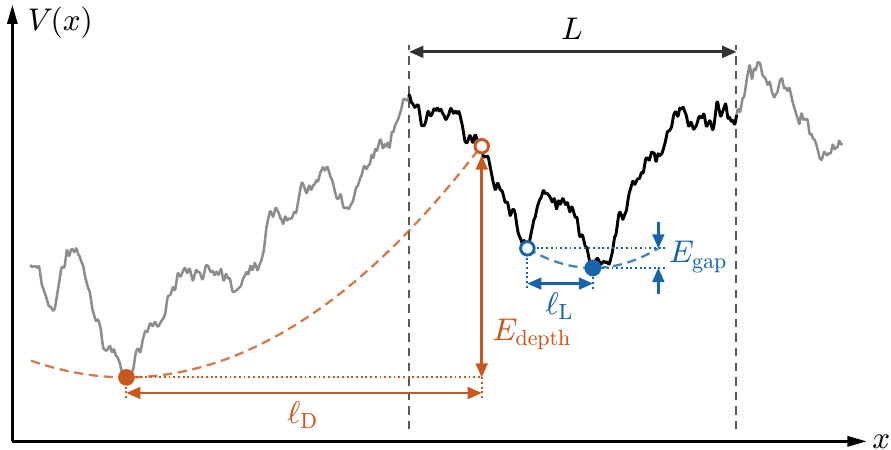}
    \caption{Gap- versus depth-Larkin balance, sketched on a one-dimensional rough potential $V(x)$ with $H=3/4$. The dashed vertical lines delimit the window of width $L$ supporting the data (black), the grey curve being a hypothetical continuation of the disorder beyond it for a softly constrained potential. The data state (filled blue dot) and its closest competitor (open blue circle) are a distance $\lL$ apart and differ by $\Egap$. From a generic point of the landscape (open orange circle), the deepest state within reach (filled orange dot) lies $E_{\rm depth}$ lower, at a distance $\lD$. The dashed parabolas illustrate the balances $\kappa\lL^2\sim\Egap$ and $\kappa\lD^2\sim E_{\rm depth}$ (note that here their stiffnesses are not equal in this one-dimensional sketch, which is only illustrative). As the depth greatly exceeds the gap, $E_{\rm depth}\sim N\Egap$ in high dimension, the depth-Larkin length reaches beyond the data window.}
    \label{fig:gap-depth}
\end{figure}

Note that one could also balance the confinement against the depth of the \textit{deepest} state within distance $\ell$, $E_{\rm depth}(\ell)\sim\sigma\ell^H\sqrt N=N\,\Egap(\ell)$, which is enhanced by the extreme-value factor $\sqrt{2\log\mathcal M}\sim\sqrt N$ instead of compressed by it. This leads to what we call the depth-Larkin balance, $\kappa\lD^2\sim E_{\rm depth}(\lD)$, which gives
\begin{equation}
  \lD\sim\left(\frac{\sigma\sqrt N}{\kappa}\right)^{\frac1{2-H}},
  \label{eq:depth-larkin}
\end{equation}
i.e. $\lD(\kappa)=\lL(\kappa/N)$. In diffusion, this scale is never realised, however. Indeed, the well is centred on a noised data point, which already sits in a low-lying state of a measure supported on the sphere, so that moving away from it gains only a gap, not a depth, as the posterior free-energy difference~\eqref{eq:log-odds} of Sec.~\ref{sec:attribution} makes explicit. At $\kappa=\tkappa/N$ the depth balance would moreover require $\lD\gg L\sqrt N$, beyond the size of the data sphere, see Fig.~\ref{fig:gap-depth} for an illustration. The depth-Larkin scale is instead the relevant one when $\kappa$ is a genuine confining potential centred on a generic point, as for a particle or a manifold pinned in a random potential in a box of linear size $L \sqrt{N} \gg \lD$, where it reproduces the replica results~\cite{mezard1992manifolds,engle1993replicas} and underlies the cusp analysis of \citet{le2008cusps} (Sec.~\ref{sec:shocks}). As we show in Sec.~\ref{sec:foam}, the distinction between the two balances is essential in determining the shape of the cells.

\subsection{The speciation window: Descending the Parisi tree}
\label{sec:cursor}

Replacing the expression for the energy gap in Eq.~\eqref{eq:larkin-balance}, we finally obtain
\begin{equation}
  \lL(\kappa)\sim\left(\frac{\sigma}{\kappa\sqrt N}\right)^{\frac1{2-H}}=\sqrt N\left(\frac{\tsigma}{\tkappa}\right)^{\frac1{2-H}},
  \label{eq:larkin-scale}
\end{equation}
i.e. $\DL=\lL/\sqrt{N}\sim(\tsigma/\tkappa)^{1/(2-H)}$. 
For a given $\kappa$, pure states at distances smaller than 
$\lL(\kappa)$ haven't speciated whereas the ones beyond it have, and the ones at scale $\lL(\kappa)$ are speciating. 

The cascade then starts when $\lL=\ell_0$ and ends when $\lL=\ell_1$, leading to an attribution window
\begin{equation}
  \tkappa_s\sim\tsigma D_0^{-(2-H)}\;\lesssim\;\tkappa\;\lesssim\;\tkappa_c\sim\tsigma D_1^{-(2-H)},
  \label{eq:window}
\end{equation}
so that we recover the spectral speciation result $\tkappa_s\propto\tsigma$, as $D_0$ is independent of the disorder strength. This result also adds a time characterising the end of the cascade, $\tkappa_c\propto\tsigma^{2/H}\propto D_1^{-2}$, and therefore the speciation \textit{window} span is given by
\begin{equation}
  \frac{\tkappa_c}{\tkappa_s}=\left(\frac{D_0}{D_1}\right)^{2-H}\propto\tsigma^{(2-H)/H},
\end{equation}
which becomes infinitely broad when $\tsigma \to \infty$. We also remark that the window widens dramatically as $H\to 0$.

Beyond the first speciation, the Larkin scale provides a natural way of locating the resolution scale of the smoothed landscape within the Parisi tree. As long as $\lL$ lies in the continuous fRSB range, we define the corresponding Parisi level $u^\star(\kappa)$ through
\begin{equation}
  D\big(u^\star(\kappa)\big)=\frac{\lL(\kappa)}{\sqrt{N}}.
\end{equation}
Substituting the Larkin scale in the expression for $D(u)$ gives, at the level of scaling, $u^\star\sim(\tkappa^H/\tsigma^2)^{1/(2-H)}$ (note that $u$ has dimension of $\tsigma$) or, normalising by the onset where $u^\star=u_0$,
\begin{equation}
  \frac{u^\star(\tkappa)}{u_0}=\left(\frac{\tkappa}{\tkappa_s}\right)^{\frac H{2-H}},\qquad \tkappa_s\le\tkappa\le\tkappa_c .
  \label{eq:cursor}
\end{equation}
At the first speciation, $\lL=\ell_0$ and therefore $u^\star=u_0$. Note that $u_0\propto1/\tsigma$ is genuinely small at strong disorder, so that the cursor then spans a wide range, from $u_0\ll1$ for ${\tkappa}={\tkappa_s}$ to $u_1=O(1)$ for ${\tkappa}={\tkappa_c}$. As $\tkappa$ increases, $\lL$ decreases and $u^\star$ moves monotonically through the continuous branch towards $u_1=(2-H)/2$. 

\begin{figure}[t]
    \centering
    \includegraphics[width=\linewidth]{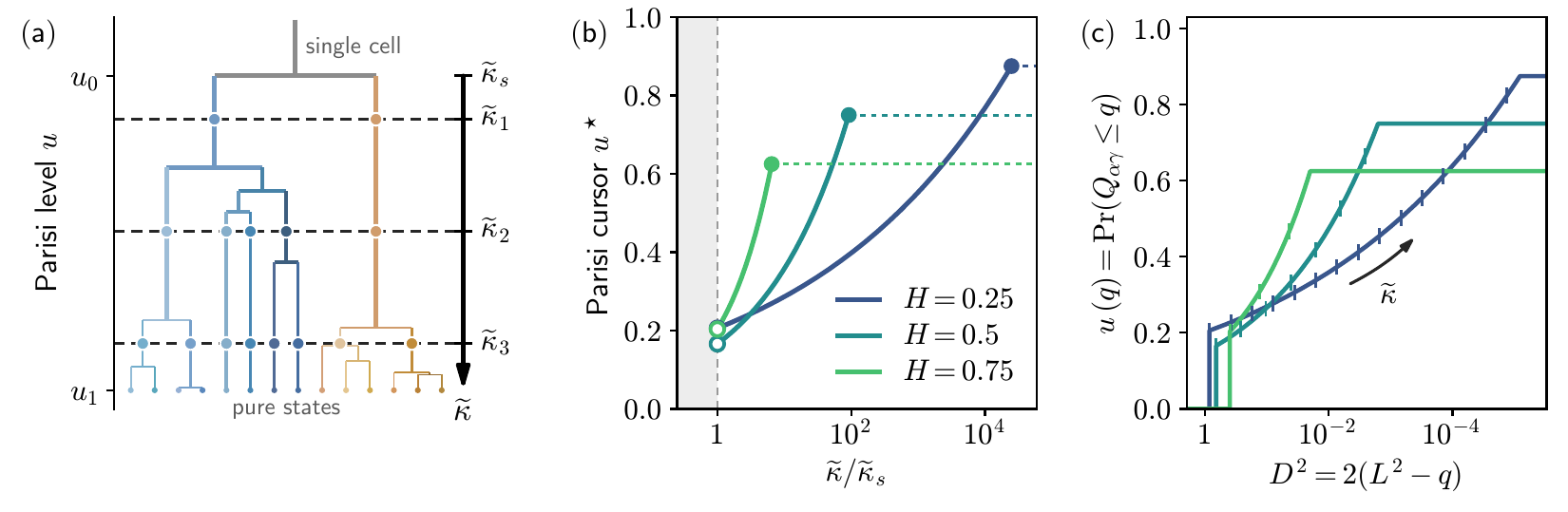}
    \caption{(a) Schematic ultrametric (Parisi) tree of the pure states of $P_0$, going down in the Parisi level $u$ from the first speciation at $u_0$ to $u_1=(2-H)/2$, where the leaves are the pure states. For $\tkappa<\tkappa_s$ the smoothed landscape is a single effective cell (grey trunk). At a given $\tkappa$, the cursor $u^\star(\tkappa)$ of Eq.~\eqref{eq:cursor} cuts the tree at the level resolved by the Larkin scale (dashed lines, with a dot on each resolved branch), and moves towards $u_1$ as $\tkappa$ increases. (b) Parisi cursor $u^\star$ versus $\tkappa/\tkappa_s$ for $H=0.25,0.5,0.75$ at $\tsigma=10$ ($L=1$). Open circles mark $u_0$ at the onset and filled circles $u_1$, reached at $\tkappa_c$, beyond which the cursor saturates (dashed). The shaded region is the single cell, $\tkappa<\tkappa_s$. (c) The same cursor on the equilibrium Parisi function $u(q)$ versus $D^2=2(L^2-q)$ (inverted axis), with ticks at log-spaced $\tkappa$ moving from $D_0$ towards $D_1$.}
    \label{fig:Parisi-tree}
    \label{fig:cursor}
\end{figure}

The smoothing thus progressively resolves finer levels of the Parisi tree, with $u^\star(\kappa)$ providing a natural parametrisation of this hierarchical cascade. Once $u^\star$ reaches $u_1$, the scale-matching construction has exhausted the hierarchy of the equilibrium data measure. This evolution is shown in Fig.~\ref{fig:Parisi-tree}(a), while the quantitative evolution of $u^\star$ and the associated $u(q)$ are shown in Fig.~\ref{fig:cursor}(b,c).

This cursor also has a direct interpretation. Two states drawn from $P_0$ lie in different cells of the foam when their distance exceeds the cell scale, $D>\DL$, which happens with probability $u$ evaluated at that scale. Averaged over the disorder, $u^\star(\tkappa)$ is therefore the probability that two states drawn from the data are already distinguished by the smoothed landscape at $\tkappa$. It grows from $u_0$ at the onset to $u_1$ at the end of the cascade.

\subsection{Geometry of the foam and shape of the cells}
\label{sec:foam}

Let us now describe more precisely the geometry of the smoothed potential in this intermediate regime. Within the rigid-state picture of Sec.~\ref{sec:larkin}, we recall that each pure state contributes to the partition function $Z(\by)$ through its cell potential $F_\alpha(\by)=F_\alpha+\frac\kappa2|\by-\bm m^\alpha|^2$,
\begin{equation}
  W(\by)\simeq-\log\sum_\alpha\e^{-F_\alpha(\by)} .
  \label{eq:cells}
\end{equation}
For a given $\by$, wherever one of the cell potentials lies below all the others by more than a few units, the sum is dominated by a single term $\alpha=\alpha^\star(\by)$, and $W(\by)\simeq F_\alpha(\by)$ is simply a harmonic well of curvature $\kappa$ centred on $\bm m^\alpha$. We call the region of the $\by$ sphere where this is the case the cell of the state $\alpha$. The boundary between two neighbouring cells is then a hyperplane perpendicular to $\bm d=\bm m^\alpha-\bm m^\gamma$, displaced from the bisector by $(F_\alpha-F_\gamma)/(\kappa\ell)$ towards the state of higher free energy, whose cell is accordingly smaller. The cells therefore tile $\by$-space as a weighted Voronoi, or more precisely a power (Laguerre) diagram of the state centres as illustrated in Figs.~\ref{fig:two-sphere-sketch} and \ref{fig:eye-candy}.

At a given $\kappa$, neighbouring states are at a distance $|\bm m^\alpha-\bm m^\gamma|= \ell=\lL$. The wall separating them is characterised by the equality of the energy $F_\alpha -\kappa \by\cdot\bm m^\alpha= F_\gamma -\kappa \by\cdot\bm m^\gamma $. Moving at a distance $r$ from this wall creates an imbalance of these energies of order $\kappa \ell r$; this is of order one when $r \sim (\kappa \ell)^{-1}$ which defines the  typical thickness of the wall.

We define the sharpness of the foam's walls as the ratio between the size $\ell$ of the cells to the thickness of the walls $(\kappa \ell)^{-1}$:
\begin{equation}
\Gam(\kappa):=\kappa\lL^2=\tkappa\DL^2\propto \tkappa^{-H/(2-H)} .
  \label{eq:sharpness}
\end{equation}
As a result, walls between coarser branches, resolved earlier, are the sharpest. In fact $\Gam(\kappa)$ characterises the least sharp walls among all previously created shocks. Indeed, as mentioned in Sec.~\ref{sec:cursor}, $u_0\propto1/\tsigma$ is genuinely small at strong disorder, so that the initial foam has $\Gam(\kappa_s) \sim1/u_0\gg1$. The sharpness then decreases along the cascade, down to $\Gam(\kappa_c) \sim1/u_1=O(1)$, where the walls become as thick as the cells and the foam picture ends. By the Larkin balance~\eqref{eq:larkin-balance}, $\Gam$ is also the free-energy gap $\Egap(\lL)$ at the cell scale, so that the foam is sharp precisely where the hierarchy is strongly condensed, i.e. where the gaps are large and $u$ is small. For the diffusion itself ($\beta=1$), the walls are not genuine discontinuities of $\nabla W$, their width being set by the noise, but they appear as such on the scale of the cells whenever $\Gam\gg1$.

Note that at the centre of a wall separating two states of equal weight, the curvature of $W$ normal to the wall is reduced to
\begin{equation}
  \bn^\top\nabla\nabla W\,\bn\le\kappa\left(1-\frac{\kappa\ell^2}4\right) = \kappa\left(1-\frac{\Gam}4\right)  ,
  \label{eq:wall-curvature}
\end{equation}
with $\bn$ the wall-normal unit vector. The above is an equality for rigid states, the fluctuations inside the states only lowering it further. Walls thus become concave, i.e. ridges of $W$ separating two basins, as soon as $\Gam>4$. While we focus on the walls separating two states here, the ultrametricity of the potential means that a cluster in fact splits into several equidistant sub-clusters. When $K$ of them have comparable free energies the walls between them meet on junctions where $K$ cell potentials coincide, i.e. on sets of codimension $K-1$. These junctions generalise the walls to higher codimension, that we shall come back to in Sec.~\ref{sec:source}.

The free-energy balance also fixes the shape of the cells, which in high dimension need not be spheroidal. By Eq.~\eqref{eq:two-state}, the free-energy difference between two cells is proportional to the normal distance $r$ to the wall between them, $|F_\gamma(\by)-F_\alpha(\by)|=\kappa\ell r$. A noised point thus lies within normal distance $r$ of a wall when the smallest free-energy difference to a competitor is below $\kappa\ell r$. Let $\rho_E(\Delta E)$ be the probability density of the smallest free-energy difference $\Delta E$ between the state of the noised point and its competitors at scale $\ell$. Assuming this density is finite and smooth as $\Delta E\to0$, the probability to lie within a small normal distance $r$ of a wall is $\rho_E(0)\,\kappa\ell r$. Since the differences are spread over a range of order $\Egap(\ell)$, $\rho_E(0)\sim1/\Egap(\ell)$, and the density of walls per unit length along the normal, i.e. the surface-to-volume ratio of the cells weighted by the noised points, is
\begin{equation}
  S_V\sim\frac{\kappa\ell}{\Egap(\ell)}=\frac1\ell\,\frac{\kappa\ell^2}{\Egap(\ell)} .
  \label{eq:surface-volume}
\end{equation}
At the gap-Larkin scale, $\kappa\lL^2\sim\Egap(\lL)$, one thus gets a surface to volume ratio $S_V\sim1/\lL$. This result indicates that the cells are very anisotropic. Indeed, a ball of radius $\ell$ in $N$ dimensions has $S_V=N/\ell$, and more generally a factor $N$ is the signature of a cell with extension similar in all directions. For instance, splitting a simplex into pyramids with apex at an interior point, one per facet, gives exactly a volume $\frac1N\sum_kh_kA_k$, with $A_k$ the area of facet $k$ and $h_k$ its normal distance to the apex, so that $S_V=N/\bar h$, with $\bar h$ the area-weighted mean distance to the facets. In order to get a surface to volume ratio of order $1/\ell$, one needs that the cells be \textit{slabs}, closed by $O(1)$ walls within reach of a noised point and open in all other directions.


The slab geometry can equivalently be seen by ranking the competitors by their free-energy difference, which grows only logarithmically  with the rank for extreme-value gaps.\footnote{Near the lowest one, the free energies of the competing states at scale $\ell$ form a Poisson process whose density grows as $\e^{\Delta E/\Egap(\ell)}$ with the height $\Delta E$ above the lowest level. The $k$-th smallest difference is therefore $\Delta E^{(k)}\simeq\Egap(\ell)\ln k$. Since a wall lies within normal distance $r$ when $\Delta E<\kappa\ell r$, the number of walls within this distance is $K(r)\simeq\exp[\kappa\ell r/\Egap(\ell)]$, i.e. $\e^{r/\lL}$ at the gap-Larkin scale.} The number of walls within normal distance $r$ then grows as $\e^{r/\lL}$, so that closing all directions (which requires at least $N$ walls) would require $r\sim (\log N) \lL$, i.e. beyond the data sphere.

Here, the depth balance $\kappa\lD^2\sim N\Egap(\lD)$ would instead give round cells, $S_V\sim N/\lD$, which is the geometry of the pinned manifolds of Ref.~\cite{le2008cusps}. Recall that the gap and depth balances involve competitors spaced by the same $\Egap$, and differ only through the penalty at the Larkin scale, which is larger by a factor $N$ for the depth balance. The same ranking then gives $\e^{Nr/\lD}$ walls within normal distance $r$, so that the cell is closed in every direction already at $r\sim\lD\log N/N$, and the exponentially many walls needed for it to be closed at comparable distances in all directions are reached at $r\sim\lD$. The cell is then nearly round, and a step of random direction leaves it after a chord $\sim\lD/\sqrt N=O(1)$.

A final geometric information can be obtained in the Hopf--Lax limit, where the curvatures of $W$ away from the walls follow from those of $V$ at the minimiser through an exact map that we derive and interpret in App.~\ref{app:foam-mobius}.

\subsection{Shocks and cusps in the score correlator}
\label{sec:shocks}

As written in Sec.~\ref{sec:diffusion}, the gradient of the smoothed potential $W$---that is the score---can be understood as the Burgers velocity field $2 \nabla W_\beta$ in the arbitrary $\beta$ generalisation. Cell (or slab) boundaries therefore have a straightforward interpretation as shocks in the $\beta \to \infty$ Hopf--Lax limit. This mapping between the replica-symmetry breaking of Gaussian random fields and Burgers shocks is at the heart of \citet{bouchaud1995scaling} and \citet{balents1996large}, where the authors consider short-range disorder, notably to understand the properties of ``Burgulence'' \cite{frisch2002burgulence, bec2007burgers}. It is also intimately related to Functional Renormalisation Group approaches to random manifolds \cite{fisher1986interface, balents1993large,le2008cusps}. In this context a natural object to consider is the velocity correlator, where shocks translate to a cusp non-analyticity at the Larkin scale. {As noted at the start of this section, however, the pinning that distinguishes the states is of order one at $\kappa=\tkappa/N$, so that the two-replica computation of this object by \citet*{le2008cusps} (LMW), carried out at $\kappa=O(1)$, does not apply directly. As we shall see, the foam nonetheless predicts its form. }

\begin{figure}[t]
    \centering
    \includegraphics{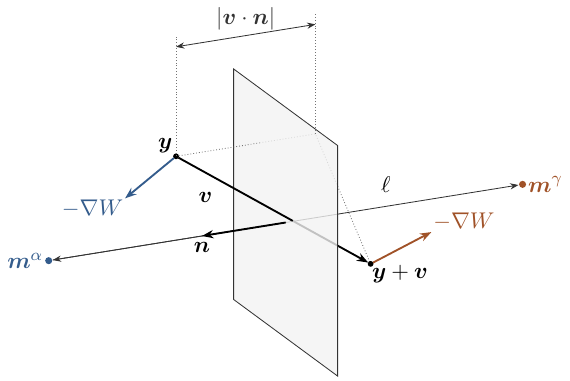}
    \caption{Wall crossing and the cusp of the score correlator. A flat wall of unit normal $\bn$ separates the cells of two states whose centres $\bm m^\alpha$ and $\bm m^\gamma$ lie a distance $\ell$ apart along $\bn$. On either side, the score $\bs=-\nabla W=\kappa(\bm m-\by)$ points towards the centre of its own state, so that across the wall $\nabla W$ jumps by $\kappa\ell\,\bn$. A step $\bv$ from the noised point $\by$ crosses the wall only if $\by$ lies within the normal distance $|\bv\cdot\bn|$ of it, which happens with probability $S_V|\bv\cdot\bn|$. For a random direction, $|\hat\bv\cdot\bn|\sim1/\sqrt N$, which gives a crossing rate $\nu\sim S_V/\sqrt N$ per unit length, each crossing adding $\kappa^2\ell^2/N$ to $\mathcal C(\bv)$.}
    \label{fig:shock-crossing}
\end{figure}

Our geometric argument for the cusp singularity of the score correlation is the following. Consider a straight step $\bv$ from a point $\by$. It crosses a wall if $\by$ lies within normal distance $|\bv\cdot\bn|$ of it, with a probability $S_V|\bv\cdot\bn|$, and for a step of random direction $\hat\bv=\bv/|\bv|$, the projection along the wall normal is typically $|\hat\bv\cdot\bn|\sim1/\sqrt N$. These two independent conditions, proximity to a wall and alignment with its normal, give a crossing rate $\nu\sim S_V/\sqrt N$ per unit length of the step. Each crossing changes the score by the jump $\kappa\ell\bn.$ For the {\it per-component} correlator of the zero-temperature potential, whose walls are genuine discontinuities, $\mathcal C(\bv)=N^{-1}\,\overline{|\nabla W_\infty(\by+\bv)-\nabla W_\infty(\by)|^2}$, the foam therefore predicts a linear cusp, with a norm that scales as the score jump squared, 
\begin{equation}
  \mathcal C(\bv)\sim \frac{S_V}{\sqrt N}\,\frac{\kappa^2\ell^2}{N}\,|\bv|,
  \label{eq:foam-cusp}
\end{equation}
which holds as long as the step crosses at most one wall. 

As mentioned above, in our generative diffusion context where $\kappa = \tkappa/N$ and the original data is on a sphere, we have cells that are in fact slabs with the surface to volume ratio $S_V \sim 1/\lL$. At the operative Larkin gap scale, we thus have 
\begin{equation}
  \mathcal C(\bv)\sim\frac{\kappa^2\lL}{N^{3/2}}\,|\bv|=\frac{\tkappa^{(3-2H)/(2-H)}\,\tsigma^{1/(2-H)}}{N^3}\,|\bv|.
  \label{eq:foam-cusp-ours}
\end{equation}

{On the other hand, if we were instead in the standard unbounded space $\kappa = O(1)$ setting of random manifolds and a generic centre, we would be in the case $S_V \sim N/\lD$, such that Eq.~\eqref{eq:foam-cusp} gives $\mathcal C(\bv)\sim\kappa^2\lD|\bv|/\sqrt N=\kappa^{(3-2H)/(2-H)}\tsigma^{1/(2-H)}|\bv|$, with \emph{no dependence on} $N$---the factor $N$ of round cells and the scale $\lD\propto\sqrt N$ compensating the factors $1/\sqrt N$ and $1/N$ of Eq.~\eqref{eq:foam-cusp}. This matches exactly the scaling of the cusp in the functional renormalisation group solution of LMW in their single-shock regime,\footnote{In this regime the step is short compared with the typical chord $\lD/\sqrt N=O(1)$ of a round cell, so that at most one wall is crossed. Longer steps, $|\bv|\sim\sqrt N$, cross many walls at all scales of the hierarchy, and the linear cusp gives way to the full functional solution of LMW.} which supports the geometric picture outlined here. The exponents in $\tkappa$ and $\tsigma$ of Eq.~\eqref{eq:foam-cusp-ours} are thus identical to those of this depth-Larkin setting, the amplitude being suppressed by $N^{-2}$ from $\kappa=\tkappa/N$ and by $N^{-1}$ from the slab geometry.


Turning the scaling argument presented here into a proper replica computation is not straightforward, for the reason given at the start of this subsection and encountered throughout this work. At $\kappa=\tkappa/N$ the pinning that distinguishes the states is of order one and does not move the extensive saddle point, so that the cusp would have to be extracted from the $O(1)$ reweighting of the tree by two planted noisy points, $\by$ and $\by+\bv$, as for the posterior over states of Sec.~\ref{sec:attribution}. We therefore leave this computation for future work.}

\section{The denoising error across scales}
\label{sec:attribution}

Generative diffusion can be seen as a succession of denoising problems. It is therefore useful to reformulate the geometric picture we have uncovered in terms of the Bayes-optimal denoiser, whose error $\Delta(\kappa)$, defined in Eq.~\eqref{eq:mmse} above, decreases from the prior width $L^2-\bar q$ to zero along the reverse process. Indeed, in the context of a trained model, the denoising error is essentially the loss that a deep network is trained against. We first study the behaviour of this denoising error across the speciation window described above, before finally computing it for $\kappa=O(1)$, i.e. when one denoises within the selected state.

\subsection{Speciation regime: the foam seen by the denoiser}
\label{sec:attribution-speciation}

In an inference context, the Larkin balance of Sec.~\ref{sec:larkin} acquires a direct reinterpretation. Consider a noised point $\by=\bm m^\alpha+\bz/\sqrt\kappa$ generated from or ``planted in'' a pure state $\alpha$, treated as rigid. The denoiser aims at finding the state of the signal, $\alpha$, and weighs each state $\gamma$ by its posterior probability $\pi_\gamma(\by,\kappa)$ .  Recalling the pure state decomposition of the measure with $w_\alpha\propto\e^{-F_\alpha}$, the posterior free-energy difference between a competitor and the planted state, i.e. their log-odds ratio, is
(App.~\ref{app:attribution-posterior})
\begin{equation}
  \log\frac{\pi_\gamma}{\pi_\alpha}=-(F_\gamma-F_\alpha)-\frac{\tkappa D^2_{\alpha\gamma}}2+\sqrt{\tkappa} D_{\alpha\gamma}\,\xi ,
  \label{eq:log-odds}
\end{equation}
where $D^2_{\alpha\gamma}=2(q_1-Q_{\alpha\gamma})$ is the squared distance between their centres and $\xi\sim\mathcal N(0,1)$. This is the Larkin balance~\eqref{eq:larkin-balance} seen from the planted state, with three terms of different origins. The gap $F_\gamma-F_\alpha$ is the prior free-energy difference between the two states. The penalty $\tkappa D^2_{\alpha\gamma}/2$ is the cost of moving away from  state $\alpha$. The tilt $\sqrt{\tkappa} D_{\alpha\gamma}\,\xi $, of zero mean and independent of the disorder, is due to the displacement of $\by$ with respect to $ \bm m ^\alpha$ and 
 is small compared with the penalty for $\tkappa D^2_{\alpha\gamma}\gg1$. For two branches of equal weight, $F_\gamma=F_\alpha$, Eq.~\eqref{eq:log-odds} reduces to the speciation criterion of \citet{achilli2026theory}, the penalty being compared with its fluctuations.
 
 In this reading, a wall of the foam is where the denoiser attributes equal probability to two states. Indeed, if the planted state $\alpha$ has lower free energy, then $\pi_\alpha>\pi_\gamma$ at all times (i.e. all $\tilde\kappa$). If on the contrary $F_\alpha>F_\gamma$, so that the planted configuration is in the ``minority branch'' of the split, the denoiser finds the right $\alpha$ provided that $\tilde\kappa> 2(F_\alpha-F_\gamma)/ D_{\alpha\gamma}^2$. Therefore the attribution of the right label by the denoiser follows the Larkin cascade.

Within the rigid state approximation relevant for the speciation regime, the denoising error can be split as
\begin{equation}
  \Delta\simeq\Delta_1+\Delta_c(\tkappa),
\end{equation}
where $\Delta_1$ corresponds to the error due to intra-state fluctuations of the draw, while
\begin{equation}
\Delta_c=\frac1N\,\mathbb E\left|\bm m^\alpha-\sum_\gamma\pi_\gamma\bm m^\gamma\right|^2=q_1-\mathbb E\sum_\gamma\pi_\gamma Q_{\alpha\gamma} ,
\label{eq:centre-mmse}
\end{equation}
where $\mathbb E$ denotes the average over the random potential, over the state $\alpha$ drawn with its weight $w_\alpha$ and over the noise $\bz$, i.e. over the draw of $\by$. The second form follows from expanding the square, using $|\bm m^\gamma|^2/N=q_1$ for all states, and exchanging $\alpha$ with a posterior sample i.e. enforcing the Nishimori identity of Sec.~\ref{sec:diffusion}. As we increase $\tkappa$ starting from 0, $\Delta_c$ decreases from $\Delta_c(0)=q_1-\bar q=\Delta_m$, the interstate width of Eq.~\eqref{eq:Delta-m}, to $\Delta_c = 0$ at the end of the attribution.

\begin{figure}
    \centering
    \includegraphics[width=\linewidth]{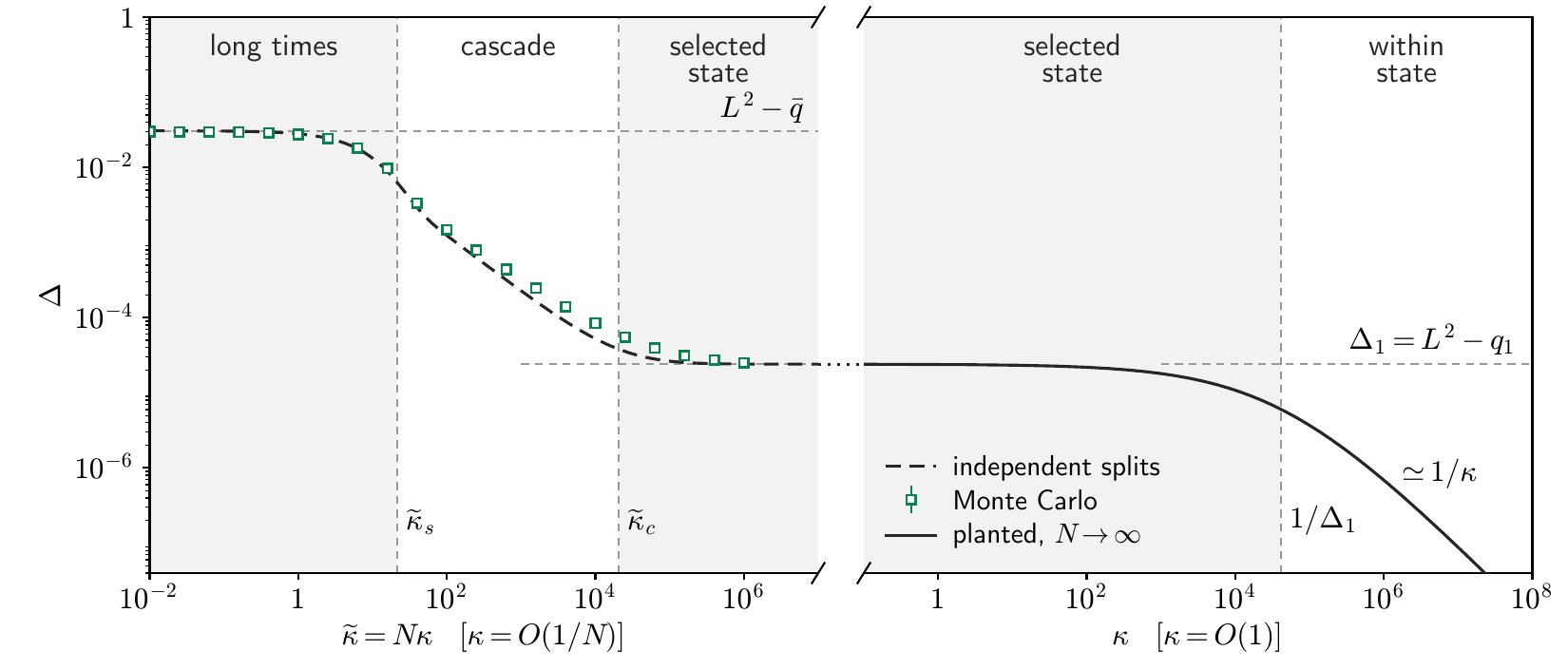}
    \caption{Per-coordinate denoising error $\Delta$, averaged over the disorder, over the whole reverse schedule, on the two scales of $\kappa$ joined at the plateau $\Delta=\Done$ (dotted across the cut); $H=1/2$, $\tsigma=24$. Left, $\kappa=O(1/N)$, against $\tkappa=N\kappa$: squares, Monte Carlo of the rigid-state posterior over states on a sampled Parisi tree (App.~\ref{app:mc}); dashed, independent splits, Eq.~\eqref{eq:independent-splits}. Right, $\kappa=O(1)$: planted branch, Eq.~\eqref{eq:planted-LR}.}
    \label{fig:two-scale}
\end{figure}

We can go further and describe the drop in $\Delta_c$ by assuming that the splits met along the branch of the true state in the tree (Fig.~\ref{fig:Parisi-tree}(a)) are independent. At a level $q$ of the tree, the log-odds given by Eq.~\eqref{eq:log-odds} shift the free-energy difference $f=F_\gamma-F_\alpha$ between a competing branch and the branch of the planted state by $\delta f=c-\sqrt{2c}\,\xi$ in favour of the latter, with $c=\tkappa(q_1-q)$. For $u(q)\ll1$, the two branches are typically very unequal, their difference being set by the gap of Eq.~\eqref{eq:gap-x}, $f\sim\Egap\sim1/u(q)\gg1$, and the denoiser picks the wrong branch only at the rare balanced splits. Averaging over these, the probability of a wrong guess is reduced by a factor $G(c)$ compared with a guess drawn from the prior, which decreases from $G(0)=1$ to $G(c)\propto c^{-1/2}\e^{-c/4}$ (App.~\ref{app:splits}). A split is thus decided once $\tkappa(q_1-q)=O(1)$. Weighting each level by the probability $P(q)=\dd u/\dd q$ that a prior guess leaves the branch of the true state there, and by the cost $q_1-q$ of such a guess, gives
\begin{equation}
  \Delta_c(\tkappa)\simeq\int\dd q\,P(q)\,(q_1-q)\,G\big(\tkappa(q_1-q)\big).
  \label{eq:independent-splits}
\end{equation}
We plot the total error $\Delta$ in the regime $\kappa = \tkappa/N$ in Fig.~\ref{fig:two-scale}. At $\tkappa=0$, $\Delta_c$ reduces to $q_1-\bar q$, and as $\tkappa$ grows the splits are decided one after the other, the coarsest first, consistent with our foam picture. The drop thus sets in when the coarsest split is decided, $\tkappa(q_1-q_0)=O(1)$, i.e. at $\tkappa\sim1/D^2_0$, independently of $\tsigma$. Inside the window, only the levels with $D^2\lesssim1/\tkappa$ are still undecided, and they keep their prior cost, so that
\begin{equation}
  \Delta_c(\tkappa)\sim D^2\,u(D)\big|_{D=1/\sqrt{\tkappa}}\propto\tsigma^{-1}\tkappa^{-(2-H)/2}.
  \label{eq:Dc-window}
\end{equation}
As in the Larkin cascade, $\tsigma$ and $H$ enter only through $u$. This decay reaches the width of the states, $\Delta_c\sim\Done$, at $\tkappa\sim1/D_1^2\propto\tsigma^{2/H}$, the end $\tkappa_c$ of the speciation window, where $u=O(1)$. Beyond it the attribution appears complete, still at $\kappa=o(1)$, which we check in Sec.~\ref{sec:short}.

This approximation is expected to hold when balanced splits are rare along the branch, $u\ll1$, i.e. at strong disorder, and to become less precise towards the end of the window, where $u\to u_1=1-H/2$.  
A direct Monte Carlo sampling of the posterior on random trees of states with the statistics of \citet{mezard1985microstructure} (App.~\ref{app:mc}), which does not include such approximations, proves to follow Eq.~\eqref{eq:independent-splits} closely, with deviations that grow only towards the end of the window (Fig.~\ref{fig:two-scale}), as expected. 

Note that, being an averaged quantity, $\Delta_c$ need not follow the typical case described in Sec.~\ref{sec:larkin}, in which the planted state lies in the dominant branch of every split and is found from the outset. It is instead dominated by the rare draws in which one of these splits is balanced, which leads it to start dropping at $\tkappa=O(1)$, independently of $\tsigma$, whereas the typical speciation begins later at $\tkappa_s\propto\tsigma$ (see Fig.~\ref{fig:two-scale}).

\subsection{Within the selected state: a replica-symmetric posterior}
\label{sec:short}\label{sec:planted}

Once the cascade has reached the finest level of the hierarchy, at $\tkappa_c$, what remains at $\kappa=O(1)$ is denoising \emph{within} the selected state, whose error $\Delta(\kappa)$ decreases from the width $\Done$ of the state to 0 as $\kappa\to\infty$. The pinning term $\kappa\,\by\cdot\bx$ is then extensive, and a replica computation applies. Since $\by=\bx_0+\bz/\sqrt\kappa$ is generated from the data, the relevant object is the quenched, trajectory-averaged log-partition function of the internal measure from a \textit{planted} configuration,
\begin{equation}
  \mathcal F(\kappa)=\overline{\frac1{\mathcal Z}\int\dd\bx_0\,\e^{-V(\bx_0)}\,\mathbb E_{\bz}\log Z(\by,\kappa)},
  \qquad \by=\bx_0+\frac{\bz}{\sqrt\kappa},
  \label{eq:planted-object}
\end{equation}
with $\mathcal Z=\int\dd\bx\,\e^{-V(\bx)}$. The noise couples the planted configuration $\bx_0$ and the replicas of the internal measure symmetrically, $\bx_0$ being itself a posterior sample. This is the replica-level expression of the Nishimori identity, which makes the computation tractable (App.~\ref{app:planted-object}). We look for a saddle point in which the replicas of $P_0$ keep their equilibrium fRSB structure, while those of the internal measure follow the branch of $\bx_0$ with a single mutual overlap $q_\post\geq q_1$, i.e. a replica-symmetric posterior of width $\Delta=L^2-q_\post\leq\Done$. The lower plateau and the whole continuum of the Parisi solution then drop out, and the plant enters only through $\Done$ (App.~\ref{app:planted-evaluation}). With the within-state ``stiffness''
\begin{equation}
\mu(\Delta):=\frac1\Delta + 2\cB'(2\Delta),
\end{equation}
the single saddle-point equation reduces to
\begin{equation}
  \mu(\Delta)=\mu(\Delta_1)+\kappa,
  \label{eq:planted-saddle}
\end{equation}
i.e. pinning raises the within-state stiffness by exactly $\kappa$. For our long-range family, with $\tilde\Delta=\Delta/\Done$,
\begin{equation}
  \kappa\Delta_1=\frac1{\tilde\Delta}-1-\frac{\tilde\Delta^{H-1}-1}{1-H},
  \label{eq:planted-LR}
\end{equation}
which for $H=\tfrac12$ gives $\Delta(\kappa)=\Done/(1+\sqrt{\kappa\Done})^2$. The error thus depends on $\kappa$ only through $\kappa\Done$, and crosses over from $\Done$ for $\kappa\Done\ll1$ to $\Delta\simeq1/\kappa$ for $\kappa\Done\gg1$. This error is shown on the right of Fig.~\ref{fig:two-scale}, and for several values of $H$ in App.~\ref{app:planted-stability}.

Beyond the denoising error itself, this computation provides an important consistency check of our picture. If a single state has been selected, the overlap between two samples of the posterior at fixed $\by$ takes a single value, i.e. the posterior is replica symmetric, whereas several competing states would break replica symmetry. Here, the replica-symmetric solution is stable as long as its replicon eigenvalue
\begin{equation}
  \lambda_R^{\post}=\frac1{\Delta^2}-4\cB''(2\Delta)=\frac1{\Delta^2}\left[1-\left(\frac\Delta{\Delta_1}\right)^H\right],
  \label{eq:planted-replicon}
\end{equation}
is positive, where the second form uses Eq.~\eqref{eq:q1}. It is therefore positive for every $0<\Delta<\Done$ and vanishes only at the equilibrium endpoint $\Delta=\Done$, where the pure state is marginal, confirming that the posterior is indeed replica symmetric at any fixed $\kappa>0$, and a single state has been selected, consistently with the speciation picture. Note that in this check, planting $\by$ is essential. Taking a random $\by$ independent of the disorder, one instead finds a fRSB internal measure up to a finite $\kappa$ (App.~\ref{app:planted-independent}), i.e. a competition between pure states persisting far beyond the speciation regime, which would be a contradiction. 

As $\kappa\to0^+$, the planted branch tends to the width $\Done$ of a single pure state, not to the prior error $L^2-\bar q$, while its stability margin closes, $\lambda_R^\post\to0$. The limits $N\to\infty$ and $\kappa\to0$ thus do not commute, the posterior recovering the whole tree only in the speciation regime $\kappa=\tkappa/N$ that is invisible to this replica computation and required the Larkin arguments developed above.

\section{Beyond the exact score: imperfect reverse dynamics}
\label{sec:source}

The focus of this work has been the exact score, which one could hope to learn given a perfectly adapted architecture and enough data. In practice however, for any finite number of samples, the learned score is only an approximation of it. In the context of long-range correlated data, for which we have shown that the reverse process gradually uncovers a generative foam, it appears relevant to ask how errors in the learned score interact with this spatial structure. In this section we take a first step in this direction by asking where a generic, noisy error of the score feeds probability in the exact dynamics, and leave the study of trained denoisers on structured data to a subsequent work.

Recall that in terms of the smoothed score, $\bs=-\nabla W$, the exact reverse dynamics~\eqref{eq:reverse-FP} reads $\partial_\tau P=\nabla\cdot[2\nabla W\,P]+\nabla^2P$. To avoid precisely parametrising the learned score, we propose to assume that its error is centred, isotropic and decorrelates quickly along trajectories. On average over the errors, it then simply adds an effective white noise to the drift, i.e. an extra diffusivity $\epsilon$,
\begin{equation}
  \partial_\tau P^\epsilon_\tau=\nabla\cdot\big[2\nabla W\,P^\epsilon_\tau\big]+(1+\epsilon)\nabla^2P^\epsilon_\tau.
  \label{eq:perturbed}
\end{equation}
Assuming this noise is small, we expand to the first order with $P^\epsilon=P+\epsilon \delta P$. The error part $\delta P$ then follows
\begin{equation}
\partial_\tau \delta P=\mathcal L_\kappa \delta P+P \mathcal{J}
\end{equation}
where $\mathcal L_\kappa$ is the Fokker-Planck operator of the exact reverse dynamics ($\epsilon=0$), $\mathcal L_\kappa \varphi=\nabla\cdot[2\nabla W\,\varphi]+\nabla^2\varphi$, and the amplifying ``stray'' source term is
\begin{equation}
  \mathcal{J}(\by,\kappa):=\frac{\nabla^2P}{P}=|\nabla W|^2-\nabla^2W .
  \label{eq:source-def}
\end{equation}
This source term redistributes probability without changing the total mass, $\int P \mathcal{J}=0$ (App.~\ref{app:G-exact}).

Equation~\eqref{eq:source-def} already suggests that the geometry of the foam is important, with notably the second term, $-\nabla^2W$, \emph{positive} where $W$ is concave, i.e. across the walls between the cells of the foam, and negative close to the peaks of the exact probability distribution $P$, where $W$ is locally minimal.  Although the score error is isotropic, the probability it feeds is therefore not uniform, but set by the geometry of the foam, as probability is removed from peaks, and added along walls.

\begin{figure}[t]
    \centering
    \includegraphics[width=\linewidth]{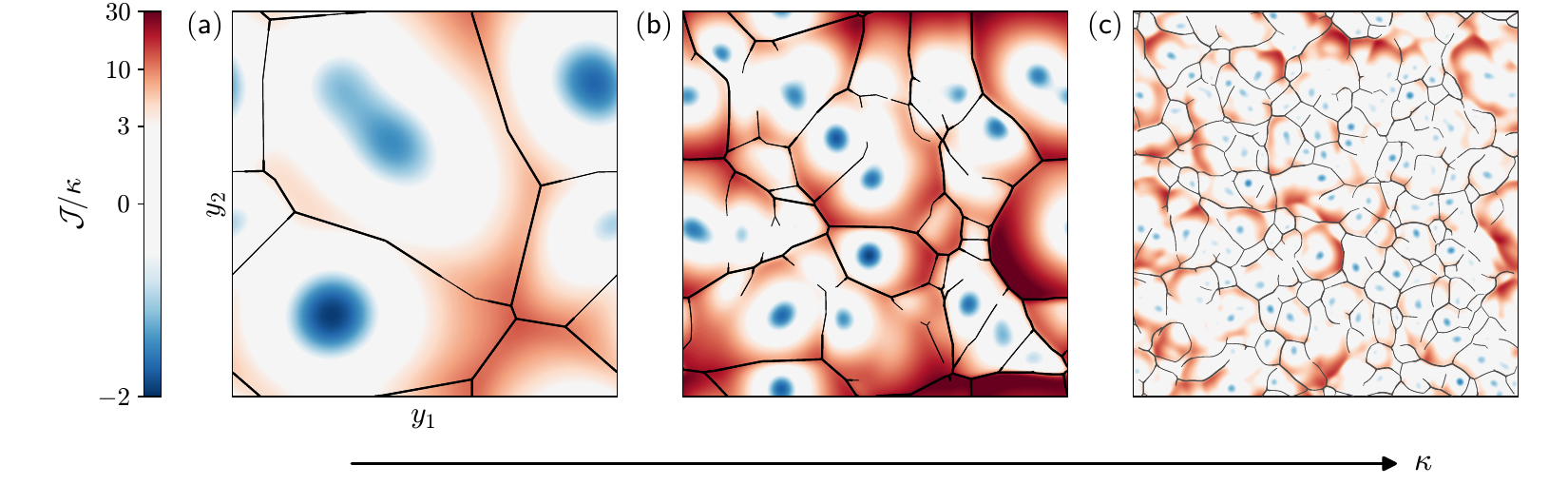}
    \caption{Source $\mathcal{J}/\kappa$, Eq.~\eqref{eq:source-def}, in the two-dimensional foam of Fig.~\ref{fig:eye-candy} (same field, unit temperature, periodic box), at three stages of speciation, with $\kappa$ increasing from (a), when only a few cells have formed, to (c), at the end of the cascade. The walls of $W$ at each time are drawn as lines. As the cells form, $\mathcal{J}$ is negative at their cores (blue) and positive across the walls (red).}
    \label{fig:foam-source}
\end{figure}

In our setting, this dependence on the geometry can be made explicit (App.~\ref{app:G-junctions}). In the sharp foam limit, $\Gam=\kappa\lL^2\gg1$, the source is of order $\mathcal{J} \sim \kappa\Gam \propto \kappa^{(2-2H)/(2-H)}$ on the walls between two cells of size $\lL$, and reaches a maximum value of the same order on the junctions where $K\sim\sqrt\Gam$ cells meet, i.e. on shocks of codimension $K-1$. Errors of the score thus feed probability not only onto the walls, but also onto regions where many states compete, and where the exact denoiser outputs a barycentre of several of them. Figure~\ref{fig:foam-source} shows this source in a two-dimensional foam at three stages of speciation. As the cells form, $\mathcal{J}$ is negative at their cores and positive across the walls that separate them, so that the geometry of the foam directly sets where errors of the score feed probability.

Note that the source in itself only sets where excess probability is fed (compared to the exact reverse scheme). Where it ends up also depends on its transport by $\mathcal L_\kappa$, and thus on the case at hand. In the illustrative two-dimensional foam of Fig.~\ref{fig:foam-source} for instance, it mostly broadens the occupied cells and shifts weight between states during speciation, a reweighting that is frozen once the walls are sharp, rather than accumulating on the walls. The late stage of the dynamics matters as well. While the states are unresolved, $\kappa\Done\ll1$, the exact drift $-2\nabla W\simeq-2\kappa(\by-\bm m^\alpha)$ restores each cell towards its centre at a rate $2\kappa$, and pulls back the excess fed into it. It is likely that a trained network does not reproduce such a strongly restoring late-time drift, so that in practice the probability fed during speciation may persist much more than in these exact dynamics that are only illustrative.

This is therefore where the foam may matter for imperfect generation. On the walls and junctions, the exact denoiser outputs barycentres of the competing states, $\xhat=\sum_\gamma\pi_\gamma\bm m^\gamma$, and in the foam these involve $K=O(1)$ states because the tree is condensed and few states are therefore balanced at any given noised point. The extent to which such interpolations are good samples, i.e. whether the potential at a barycentre is predicted by its neighbours as in so-called ``Kriging''~\cite{krige1951statistical,matheron1963principles}, and whether the mass fed onto the walls survives the subsequent dynamics of trained models, will be the subject of a forthcoming dedicated paper.

\section{Discussion}
\label{sec:discussion}

In this paper, we have studied the exact process of generative diffusion for Gaussian random fields with long-range correlations in high dimension. Such random fields lead to a hierarchical organisation of samples in space which appears to be plausibly present in real-world data such as images. What we have shown is that these correlations and hierarchy lead to a speciation \textit{cascade} on the typical speciation timescale first identified in \citet*{biroli2024dynamical}. There, the smooth potential associated to generative diffusion develops as a \textit{foam}, with cells associated to the pure states of the Gibbs measure separated by shock-like boundaries, the geometry of which we have described. At shorter times, the details of the potential within the pure states of the measure are finally resolved. The denoising error of the exact score follows these stages, its averaged drop during speciation being set by the Parisi function alone, while we showed that the walls of the foam are where generic errors of an imperfect score feed excess probability during the reverse process.

As highlighted by the combination of different methods that we resorted in, no single theory adequately describes the whole reverse process. At $\kappa=O(1)$ ($t= O(1)$) the pinning term $\kappa\,\by\cdot\bx$ is extensive, and the planted replica saddle point of Sec.~\ref{sec:short} controls the posterior. In the speciation regime $\kappa=\tkappa/N$ (meaning $t = O(N)$) it is instead of order one, an $O(1)$ perturbation of an extensive problem that a saddle point at $N\to\infty$ does not see, leading us to develop alternative scaling arguments. This difficulty can be summarised in the clear importance of the order of limits for the mean posterior overlap $\bar q_\post^{(N)}(\kappa):=N^{-1}\overline{\mathbb E\,|\xhat(\by,\kappa)|^2}$ at finite $N$, since
\begin{equation}
  \lim_{\kappa\to0}\lim_{N\to\infty}\bar q_\post^{(N)}(\kappa)=q_1,\quad \text{while} \quad
  \lim_{N\to\infty}\lim_{\kappa\to0}\bar q_\post^{(N)}(\kappa)=\bar q=\int_0^1 \dd u \, q(u)<q_1 .
  \label{eq:limits}
\end{equation}
The two limits probe opposite sides of the attribution regime $N\kappa=O(1)$. Selection among states happens at $\kappa=\tkappa/N$; resolution of configurations within the selected state happens at fixed $\kappa>0$.

A second subtlety that we have encountered, as in almost every problem related to disordered systems, is the difference between typical and average behaviour. Sample by sample, the geometric and informational notions of speciation coincide. 
Indeed, a trajectory ending in the minority branch of a split commits to it once its planted bias $\tkappa D^2/2$ overcomes the free-energy difference $f$ between the branches, which is also when this branch acquires a basin of its own in $P_t$. In a typical sample, $f$ is of the order of the gap, $\Egap(D)\sim1/u(D)\gg1$ at the coarse levels, so that speciation occurs at the Larkin balance, from $\tkappa_s\propto\tsigma$ on. Only a fraction $\sim u(D)$ of the samples have a balanced split, $f=O(1)$, yet these dominate disorder averages such as the denoising error of Fig.~\ref{fig:two-scale}, which therefore start to move at $\tkappa\sim1/D^2_0$, independently of $\tsigma$. 
The spectral criterion of Sec.~\ref{sec:speciation}, a spinodal evaluated with the disorder-averaged covariance, describes neither exactly, but nonetheless still shares the typical scaling. The two descriptions merge at the end of the cascade, $\tkappa_c$, where $u_1=O(1)$ and balanced splits are no longer rare.

Note that several features of this picture rely on the condensation of the Gibbs measure. With $M=\e^{O(N)}$ comparable states, a trajectory commits only once its planted bias $\kappa\ell^2/2$ overcomes the entropy $\log M=O(N)$ of its competitors, which pushes speciation to $\kappa=O(1)$, as shown explicitly for Gaussian mixtures with exponentially many centroids~\cite{ventura2026emergence}. This would be our situation for a truncated landscape in which exponentially many states, such as the local minima of high-dimensional random landscapes~\cite{bray2006statistics,fyodorov2007classical,ros2023high}, carry comparable weights instead of being Gibbs-weighted. In such a scenario, we would expect our picture to mostly hold, albeit displaced entirely at $\kappa = O(1)$.
{However, for $M$ growing subexponentially with $N$, we expect the scalings of the present paper to hold even if such low-lying states are uniformly weighted instead of being Gibbs weighted.}

The long-range correlations matter as well. For a short-range potential, whose Parisi function has a single step, the tree has a single level, and the cascade collapses into a single first-order speciation, with no window. In general, the walls of the foam are the shocks of Burgers turbulence with random initial conditions~\cite{sinai1992statistics,kida1979asymptotic,frachebourg2000exact,bertoin1998inviscid,bec2007burgers}, a connection we have only used at the level of scaling. 
We expect the structure of our arguments, namely the Larkin balance between pinning and free-energy gaps and the parametrisation of the cascade by the Parisi level $u^\star(\kappa)$, to carry over to other hierarchical measures, whereas the exponents are specific to the long-range Gaussian model. Whether this is the case, in particular for non-Gaussian potentials, remains to be seen. 
Finally, as mentioned above, in a variance-preserving schedule the same results should hold with $\kappa=1/(\e^{2t}-1)$, albeit without the required contraction of the noise shell towards the data sphere.

Several points remain open. In particular, the rigid-state picture is not controlled near $\tkappa_c$, where $\Gam=O(1)$, meaning that the walls are as thick as the cells, and the finite-$N$ corrections to the separation of scales have not been estimated. More broadly, the Larkin balance and the foam structure (cell size, anisotropy) remain at the level of scaling arguments. A two-copy replica computation analogous to that of~\citet{le2008cusps}, with two points $\by$ and $\by+\bv$ but at $\tkappa=O(1)$, would be necessary to characterise the geometry of the foam more precisely.

Overall, our results give a concrete sense in which following an entire reverse process matters for hierarchically structured data, as the commitments to the successive levels of the hierarchy are spread over a window of noise levels that broadens with the disorder, each being decided at its own scale. Whether the trajectories of trained models feel the walls of the foam we describe, and whether they could benefit from the spatial structure by generating interpolations between states there, will be the subject of a forthcoming work on finite data.

\subsection*{Acknowledgments}
We want to thank Y. Fyodorov and P. Le Doussal for very insightful comments, as well as B. Achilli and E. Ventura for stimulating discussions. The work of J.\,G.-B. was supported by the European Union’s Horizon Europe program under the Marie Skłodowska-Curie grant agreement No.~101210798. J.\,G.-B. also thanks École Normale Supérieure's Centre de Sciences des Données for its hospitality during the completion of this project.

\subsection*{Generative AI statement}

We acknowledge the use of large language models, primarily Anthropic's Claude Opus 4.7--5.5, which were mainly used to write numerical code and figure scripts, to check derivations and challenge scaling arguments, and to suggest edits to the manuscript. All results were checked by the authors, who take full responsibility for the content, and who wrote the final text.

\bibliographystyle{unsrtnat}
\bibliography{refs}

\clearpage
\appendix
\addtocontents{toc}{\protect\setcounter{tocdepth}{1}}

\section{Replica theory of the data distribution}\label{app:replica}

We recall here the known results for the Gibbs measure $P_0\propto\e^{-V}$, which guide our geometric understanding of the generative diffusion. We follow the replica analysis of M\'ezard \& Parisi~\cite{mezard1990interfaces,mezard1991replica}, see also \cite{engle1993replicas}, which was later refined by \citet{fyodorov2007classical}. We use the notations introduced in Sec.~\ref{sec:potential}.

\subsection{Replicated partition function and overlap action}\label{app:A-action}

Consider the Gibbs measure of the unsmoothed random potential on $\Omega=S^{N-1}(L\sqrt N)$, with partition function
\begin{equation}
  \mathcal Z_V=\int_\Omega\dd\bx\,\e^{-V(\bx)}.
  \label{eq:A-Z}
\end{equation}
The replicated partition function can be directly averaged over the disorder due to the Gaussian nature of the randomness,
\begin{equation}
  \overline{\mathcal Z_V^n}
  =\int_\Omega\prod_{a=1}^n\dd\bx_a\,
  \exp\left(\frac12\sum_{a,b=1}^nB(|\bx_a-\bx_b|)\right).
  \label{eq:A-Zn}
\end{equation}
The integrand depends on the replicas only through their overlaps $Q_{ab}$, with $Q_{aa}=L^2$, via the rescaled correlation $\cB(z)=B(0)/N-\tsigma^2z^H$ of Sec.~\ref{sec:potential}, with derivatives $\cB'(z)=-\tsigma^2Hz^{H-1}$, $\cB''(z)=\tsigma^2H(1-H)z^{H-2}$, $\cB'''(z)=-\tsigma^2H(1-H)(2-H)z^{H-3}$, and where we have used the canonical scaling $\sigma^2=\tsigma^2N^{1-H}$, Eq.~\eqref{eq:canonical}.

Taking the direct route detailed in \citet{fyodorov2007classical}---that is, avoiding the historical Gaussian ansatz and instead taking advantage of the rotationally invariant nature of the integrand---the replicated partition function can be computed purely in terms of the overlap matrix $Q$. Up to factors subexponential in $N$,
\begin{equation}
  \overline{\mathcal Z_V^n}\asymp\int_{\mathcal Q_n}\dd Q\,\exp\left[N\mathcal S_n(Q)\right],
  \label{eq:A-overlap-integral}
\end{equation}
where $\mathcal Q_n$ is the set of positive definite matrices with $Q_{aa}=L^2$, and
\begin{equation}
  \mathcal S_n(Q)=\frac12\log\det Q+\frac12\sum_{a,b=1}^n\cB\big(2(L^2-Q_{ab})\big).
  \label{eq:A-action}
\end{equation}
The first term is the configurational entropy associated with realising a given overlap matrix (the volume of $n$-tuples of vectors with Gram matrix $NQ$ scales as $(\det Q)^{(N-n-1)/2}$), while the second results from the Gaussian disorder average. The large-$N$ problem is consequently reduced to a saddle point over $Q$. Varying an off-diagonal element gives
\begin{equation}
  (Q^{-1})_{ab}=2\cB'(D^2_{ab}),\qquad a\neq b.
  \label{eq:A-overlap-saddle}
\end{equation}
This equation is the starting point for both the replica-symmetric and Parisi solutions.

\subsection{Replica-symmetric solution and AT line}\label{app:A-RS}

For the replica-symmetric (RS) ansatz $Q_{a\neq b}=q$, let $\Delta=L^2-q$. In the replica limit, $(Q^{-1})_{a\neq b}=-q/\Delta^2$. Using $\cB'(z)=-\tsigma^2Hz^{H-1}$ in \eqref{eq:A-overlap-saddle} gives
\begin{equation}
  q=2^H\tsigma^2H\Delta^{H+1}.
  \label{eq:A-RS}
\end{equation}
The eigenvalue controlling fluctuations that break replica symmetry is the replicon eigenvalue
\begin{equation}
  \lambda_R=\frac{1}{\Delta^2}-4\cB''(2\Delta).
  \label{eq:A-replicon}
\end{equation}
The RS solution is stable when $\lambda_R\geq0$. At the de Almeida--Thouless (AT) transition~\cite{de1978stability}, $2^H\tsigma_{\rm AT}^2H(1-H)\Delta_{\rm AT}^H=1$. Combining this condition with \eqref{eq:A-RS} gives $q_{\rm AT}=\Delta_{\rm AT}/(1-H)=L^2/(2-H)$, $\Delta_{\rm AT}=L^2(1-H)/(2-H)$, and thus the AT line~\eqref{eq:sigma-AT}. For $\tsigma<\tsigma_{\rm AT}$ the spherical Gibbs measure is replica symmetric; for $\tsigma>\tsigma_{\rm AT}$ the RS solution is unstable.

\subsection{Full-RSB solution}\label{app:A-fRSB}

For $\tsigma>\tsigma_{\rm AT}$ the solution breaks replica symmetry continuously. Henceforth we place ourselves in this \emph{strong disorder} regime, where the overlap matrix must be parametrised by a nondecreasing Parisi function $u(q)$. Let its continuous part extend from $q_0$ to $q_1$, and define
\begin{equation}
  \chi(q)=L^2-q_1+\int_q^{q_1}\dd p\,u(p)=\Done+\int_q^{q_1}\dd p\,u(p),
  \label{eq:A-chi}
\end{equation}
which is constant, $\chi(q)=\chi(q_0)$, for $q\leq q_0$.

Evaluating the determinant term for a Parisi matrix, $\lim_{n\to0}n^{-1}\log\det Q=\log\Done+q_0/\chi(q_0)+\int_{q_0}^{q_1}\dd q/\chi(q)$, and the disorder term with the $n\to0$ rule $n^{-1}\sum_{a\neq b}g(Q_{ab})\to-\int_0^1\dd u\,g(q(u))$, the action per replica is, with $D^2(q)=2(L^2-q)$,
\begin{equation}
  \mathcal S[u] = \frac12\Bigg[\log\Done+\frac{q_0}{\chi(q_0)}+\int_{q_0}^{q_1}\frac{\dd q}{\chi(q)} +\cB(0)-\cB(D^2_1)-2\int_{q_0}^{q_1}\dd q\,u(q)\,\cB'\big(D^2(q)\big)\Bigg].
  \label{eq:A-Parisi-functional}
\end{equation}
The physical solution is the stable extremum over nondecreasing $0\leq u(q)\leq1$.

Stationarity with respect to $u(q)$, $q_0<q<q_1$, using $\delta\chi(p)/\delta u(q)=\theta(q-p)$, gives
\begin{equation}
  -\frac{q_0}{\chi(q_0)^2}-\int_{q_0}^{q}\frac{\dd p}{\chi(p)^2}=2\cB'\big(D^2(q)\big),
  \label{eq:A-Parisi-stationarity}
\end{equation}
the Parisi-matrix form of Eq.~\eqref{eq:A-overlap-saddle} (the left-hand side is the Parisi function of $Q^{-1}$). Differentiating once in $q$ gives $\chi(q)^{-2}=4\cB''(D^2(q))$, i.e. the vanishing of the replicon \eqref{eq:A-replicon} along the whole continuous branch (marginal stability); evaluating \eqref{eq:A-Parisi-stationarity} at $q\to q_0^+$ fixes $q_0$; continuity of $\chi$ at $q_1$ fixes $q_1$. These steps give
\begin{equation}
  \chi(q)=\frac{1}{2\sqrt{\cB''\big(2(L^2-q)\big)}},\qquad q_0<q<q_1.
  \label{eq:A-chi-saddle}
\end{equation}
Differentiating with respect to $q$ (with $\chi'=-u$) gives the continuous Parisi function [cf.\ Eq.~\eqref{eq:xq}]
\begin{equation}
  u(q)=-\frac{\cB'''\big(2(L^2-q)\big)}{2\big[\cB''\big(2(L^2-q)\big)\big]^{3/2}}
  =\frac12\,\frac{2-H}{2^{H/2}\tsigma\sqrt{H(1-H)}}\,(L^2-q)^{-H/2}.
  \label{eq:A-xq}
\end{equation}
The endpoints [cf.\ Eqs.~\eqref{eq:q1}, \eqref{eq:q0}] follow from
\begin{equation}
  \Done=L^2-q_1=\frac{1}{2\sqrt{\cB''\big(2(L^2-q_1)\big)}}=\left[\frac{1}{2^H\tsigma^2H(1-H)}\right]^{1/H},
  \label{eq:A-q1}
\end{equation}
and
\begin{equation}
  q_0=-\frac{\cB'\big(2(L^2-q_0)\big)}{2\cB''\big(2(L^2-q_0)\big)}=\frac{L^2}{2-H}.
  \label{eq:A-q0}
\end{equation}
Inverting Eq.~\eqref{eq:A-xq} with $u=u(q)$ gives Eq.~\eqref{eq:Du}. The upper breakpoint is universal, $u_1=u(q_1^-)=(2-H)/2$, while the lower breakpoint is
\begin{equation}
  u_0=u(q_0^+)=\frac{2-H}{2\tsigma\sqrt{H(1-H)}}\,D_0^{-H}\ \propto\ \frac1\tsigma.
  \label{eq:A-u0}
\end{equation}
Compactly, $u(D)=u_1\,(D_1/D)^{H}$ on the continuous branch, so that $u_0=u_1(D_1/D_0)^{H}$. The branch opens exactly at the AT threshold: $D_1=D_0$ (equivalently $q_0=q_1=q_{\rm AT}$, $u_0=u_1$) if and only if $\tsigma=\tsigma_{\rm AT}$, Eq.~\eqref{eq:sigma-AT}.

Rescaling $\bx=L\bx'$ maps $\Omega$ onto $S^{N-1}(\sqrt N)$ and $\tsigma^2D^{2H}$ onto $(\tsigma L^H)^2D'^{2H}$, $D'=D/L$, and shifts $\log\det Q$ by a constant. Hence $q/L^2$, $D/L$ and the Parisi function depend on $(\tsigma,L)$ only through $\tsigma L^H$ [as in Eqs.~\eqref{eq:sigma-AT}, \eqref{eq:A-q0}--\eqref{eq:A-u0}], while the diffusion enters only through $\kappa L^2$.

\subsection{Pair-distance distribution}\label{app:A-pair}

The variable $u=u(q)\in(0,1)$ has the following probabilistic meaning~\cite{parisi1983order}: if two pure states $\alpha,\gamma$ are drawn independently from the Gibbs measure, then $\Pr[Q_{\alpha\gamma}\leq q(u)]=u$. With $D^2_{\alpha\gamma}=2(q_1-Q_{\alpha\gamma})$ the squared distance between their centres, configurations drawn from $\alpha$ and $\gamma$ are at squared distance $D^2_{\alpha\gamma}+D^2_1$, so that $\Pr[D^2_{\alpha\gamma}+D^2_1\geq D^2(u)]=u$. The continuous part of Eq.~\eqref{eq:Du} then gives the density $\widetilde P(D)\propto D^{-1-H}$ on $(D_1,D_0)$, i.e. Eq.~\eqref{eq:rho-ell}, and the two plateaux give its atoms.

\subsection{Statistics of the pure-state weights}\label{app:A-weights}

We collect standard facts on the weights $w_\alpha\propto\e^{-F_\alpha}$ of the pure states in the Parisi tree of states \cite{MPSTV84,mezard1985random,mezard1985microstructure,derrida1986solution} (mathematically, the Ruelle probability cascade\,\cite{ruelle1987mathematical,bolthausen1998ruelle,mezard1987spin,panchenko2013sherrington}), used in Secs.~\ref{sec:covariance}, \ref{sec:larkin} and \ref{sec:attribution}.
\begin{enumerate}[label=(\roman*)]
  \item \emph{Free-energy gaps.} Within a given parent cluster, the free energies $f$ of its children at level $q$, measured from the lowest one, form near the bottom a Poisson point process of intensity
  \begin{equation}
    \varrho(f)\,\dd f\;\propto\;\e^{\,u(q)f}\,\dd f .
    \label{eq:A-PPP}
  \end{equation}
  Hence the gap between the two dominant siblings is exponentially distributed with mean $1/u(q)$, and the ratio of their weights is $U^{1/u}$ with $U$ uniform on $(0,1)$. This is the $\Egap\sim 1/u$ of Eq.~\eqref{eq:gap-x} (at $T=1$).
  \item \emph{Leaf weights.} Marginally, the weights of the terminal states are Poisson--Dirichlet $\mathrm{PD}(u_1)$~\cite{MPSTV84,mezard1985random,derrida1987statistical}, $u_1=(2-H)/2$, whence
  \begin{equation}
    \overline{\sum_\alpha w_\alpha^p}=\frac{\Gamma(p-u_1)}{\Gamma(p)\Gamma(1-u_1)},\qquad p>u_1.
    \label{eq:A-PD-moments}
  \end{equation}
  In particular $\overline{\sum_\alpha w_\alpha^2}=1-u_1=H/2$, which is $O(1)$, i.e. the measure is condensed on a finite number of states.
  \item \emph{Condensation at coarse levels.} The weights of the coarsest clusters (level $q_0$) are $\mathrm{PD}(u_0)$, with $u_0\propto1/\tsigma$, Eq.~\eqref{eq:A-u0}. By (i), a typical sample has a single dominant coarse branch, the second carrying a relative weight $\sim\e^{-O(1)/u_0}$, and only a fraction $O(u_0)$ of samples has a balanced coarse split. Sample-dependent quantities sensitive to the coarse levels---the sample overlap distribution $P_V(q)$ at low $q$, the interstate covariance $\mathbf C_{\rm inter}$ of Sec.~\ref{sec:covariance}---are non-self-averaging, and their disorder means are dominated by these rare balanced samples. The same holds at any level $q$ with $u_0$ replaced by $u(q)$.
\end{enumerate}

\section{Spectral estimate of the speciation scale}\label{app:spectral}

\subsection{Typical versus annealed spectral estimate}\label{app:C-typical}

In Sec.~\ref{sec:speciation} we assumed that the top eigenvalue $c_{\max}(\mathbf C_{\rm inter})$ of a typical sample scales as its disorder average. This fails at strong disorder. The coarse clusters carry Poisson--Dirichlet weights $\mathrm{PD}(u_0)$ with $u_0\propto1/\tsigma$ (App.~\ref{app:A-weights}), so that in a typical sample the coarse split is strongly unbalanced, the minority cluster carrying a relative weight $\sim\e^{-O(1)/u_0}$. Only the fraction $O(u_0)$ of samples with a balanced coarse split has $c_{\max}\sim ND_0^2$, and becomes unstable at $\tkappa\sim1/D_0^2$, independently of $\tsigma$. These rare samples carry the mean, $O(u_0)\,ND_0^2\propto N/\tsigma$. Equations~\eqref{eq:C-cmax-inter} and~\eqref{eq:ts} are therefore annealed statements, which nonetheless agree in scaling with the Larkin balance describing typical samples.

\subsection{Intrastate susceptibility and soft modes}\label{app:C-intra}

We give the details behind the intrastate estimate of Sec.~\ref{sec:speciation}, Eqs.~\eqref{eq:C-soft-edge}--\eqref{eq:C-cmax-intra}. With the soft edge~\eqref{eq:C-soft-edge}, the first two moments of the covariance spectrum are
\begin{equation}
  \frac1N\Tr\mathbf C^p=\int_0^\infty\dd\lambda\,\frac{\rho_K(\lambda)}{\lambda^p},\qquad p=1,2.
\end{equation}
The first integral converges at the lower edge for $\theta>0$, while the second diverges for $\theta<1$. The interval $0<\theta<1$ therefore describes a finite thermal variance together with a divergent spin-glass susceptibility. The coefficient $f$ controls the density of soft modes and has dimension $[\lambda]^{-1-\theta}$. With the cutoff~\eqref{eq:C-lambda-min},
\begin{equation}
  \frac1N\Tr\mathbf C^2\simeq\int_{\lambda_{\min}}^\infty\dd\lambda\,\frac{f\lambda^\theta}{\lambda^2}\simeq\frac{f}{1-\theta}\lambda_{\min}^{\theta-1}
  =\frac{f^{2/(1+\theta)}}{1-\theta}\left(\frac{N}{1+\theta}\right)^{(1-\theta)/(1+\theta)}.
  \label{eq:C-TrC2}
\end{equation}

It remains to determine the dependence of $f$ on the disorder strength. It follows from the intrastate width fixed by the fRSB solution, Eq.~\eqref{eq:A-q1}, $N^{-1}\Tr\mathbf C=L^2-q_1=\Done=D^2_1/2$, so that the characteristic stiffness scale is
\begin{equation}
  k_*=\Done^{-1}=\frac{2}{D^2_1}=2\big[\tsigma^2H(1-H)\big]^{1/H}.
  \label{eq:C-kstar}
\end{equation}
Assuming that the stiffness spectrum has a one-scale form, $\rho_K(\lambda;\tsigma)=k_*^{-1}\widehat\rho_H(\lambda/k_*)$, $\widehat\rho_H(v)\simeq f_Hv^\theta$, we are left with
\begin{equation}
  f=f_Hk_*^{-1-\theta}\propto\tsigma^{-2(1+\theta)/H},
  \label{eq:C-f}
\end{equation}
where $f_H=O(1)$ is a dimensionless coefficient describing the shape of the rescaled edge. Substitution into Eq.~\eqref{eq:C-lambda-min} gives $c_{\max}(\mathbf C^\alpha)\simeq[Nf_H/(1+\theta)]^{1/(1+\theta)}D_1^2/2$, i.e. Eq.~\eqref{eq:C-cmax-intra}.

\section{Low-lying gaps and a real-space GREM}
\label{app:gaps}\label{app:gaps-grem}

This appendix provides an alternative derivation of the low-lying gap scaling
\begin{equation}
  \Egap(\ell) \;\sim\; \frac{\sigma\ell^H}{\sqrt N}=\tsigma\left(\frac{\ell}{\sqrt N}\right)^H
  \label{eq:B-gap}
\end{equation}
that underlies the Larkin argument of Sec.~\ref{sec:larkin} [Eq.~\eqref{eq:gap}]. The replica route, $\Egap\sim T/u_T(q)$, is also given in Sec.~\ref{sec:larkin} [Eq.~\eqref{eq:gap-x}]. Here instead we derive the gap without replicas, from a multiscale decomposition of $V$ combined with extreme-value statistics. The decomposition defines an explicitly solvable real-space GREM, which also reproduces $u\propto D^{-H}/\tsigma$, $\ell_1\propto\sqrt N\tsigma^{-1/H}$ and $\rho(\ell)\propto\ell^{-1-H}$ (see below).

Let $V(\bx)$ be a Gaussian field in $N$ dimensions with power-law increments
\begin{equation}
  \overline{\bigl(V(\bx)-V(\bx')\bigr)^2} = 2\sigma^2|\bx-\bx'|^{2H},\qquad 0 < H < 1,
  \label{eq:B-structure}
\end{equation}
i.e. $B(r)=B(0)-\sigma^2r^{2H}$, in the inertial range $\xi_\SR\ll|\bx-\bx'|\ll\xi_\LR$. We write $r_0:=\xi_\LR=L\sqrt N$ for the largest scale. The crucial property is self-similarity: over a region of size $r$, $V$ fluctuates by $\sigma r^H$, and $V(r\bx)\overset{d}{=}r^H V(\bx)$. Hence the natural energy unit at scale $r$ is $\sigma r^H$.

Let $\bx^\star = \arg\min V$. We ask for the typical energy gap $\Egap(\ell)$ between $V(\bx^\star)$ and $V(\bx^{\star\star})$, where $\bx^{\star\star}$ is a secondary low-lying minimum at Euclidean distance $\sim\ell$ from $\bx^\star$.

In $N$ dimensions, one can place $\sim (r_0/\xi_\SR)^N$ points mutually separated by $\gtrsim\xi_\SR$, suggesting a REM count~\cite{derrida1981random} of $\sim r_0^N$ competing minima. This is wrong: long-range correlations mean that nearby points sit in the same basin. The correct object is not the number of cells but the number of \emph{independent scales} of $V$, analogous to the law of the iterated logarithm. What enters the gap at scale $r$ is an exponential number $\e^{cN}$ of candidates \emph{per scale} (the siblings of a common parent), not per cell.

Write $V$ as a sum over dyadic scales:
\begin{equation}
  V(\bx) = \sum_{k=0}^K V_k(\bx),
  \label{eq:B-decomp}
\end{equation}
with $K = \log_2(r_0/\xi_\SR)$, and $V_k$ decorrelating on the length $r_k = r_0\, 2^{-k}$ and having amplitude
\begin{equation}
  a_k = \sigma r_k^H = \sigma r_0^H\, 2^{-kH}.
  \label{eq:B-ak}
\end{equation}
The variances $a_k^2$ sum, for $H > 0$, to a finite total $\sigma^2 r_0^{2H}/(1-2^{-2H})$, reproducing $\overline{V^2}$ up to boundary corrections.

The decomposition naturally defines a tree. Partition $\Omega$ hierarchically into cells: generation-$k$ cells ($k=0,\dots,K$) have linear size $r_k$; each generation-$k$ cell is parent to $2^N$ generation-$(k+1)$ children, and the $2^N$ generation-$0$ cells are the children of the root $\Omega$. This defines a $2^N$-ary tree of depth $K$, whose leaves correspond to configurations resolved at the UV cutoff. Attach to each generation-$k$ cell an independent centred Gaussian increment of variance $a_k^2$. The value of $V$ at leaf $\alpha$ is
\begin{equation}
  V_\alpha \;=\; \sum_{k=0}^K V_k^{(\alpha)},
  \label{eq:B-leaf}
\end{equation}
with $V_k^{(\alpha)}$ the generation-$k$ increment on $\alpha$'s ancestor chain. Two leaves $\alpha, \gamma$ whose last common ancestor sits in generation $k^\star_{\alpha\gamma}$ (with $k^\star=-1$ for the root) are at distance $\ell_{\alpha\gamma}\sim r_{k^\star+1}$ and have
\begin{equation}
  \overline{(V_\alpha - V_\gamma)^2} = 2\sum_{k>k^\star_{\alpha\gamma}} a_k^2
  = \frac{2\sigma^2 r_{k^\star+1}^{2H}}{1-2^{-2H}}\sim 2\sigma^2\ell_{\alpha\gamma}^{2H},
  \label{eq:B-tree-cov}
\end{equation}
reproducing the continuum structure function \eqref{eq:B-structure}. The tree is thus a legitimate discretisation of $V$, equivalent to Derrida--Gardner's Generalised Random Energy Model (GREM)~\cite{derrida1986solution} with real-space embedding---i.e. a \emph{real-space} GREM in which leaves of the tree correspond to points of the physical domain. Ultrametricity is exact by construction, since leaves at tree-distance $d$ have covariance depending only on $d$, not on their specific identities. This ultrametricity, which in the continuum field is emergent from the RSB solution, is built-in in the dyadic version.

On this tree, $V$ is a branching random walk (BRW) with $K$ levels and step variances $a_k^2$. Greedy minimisation of the BRW gives
\begin{equation}
  V_{\min} \approx -\sqrt{2N\log 2}\sum_{k=0}^K a_k
  = -\sigma r_0^H\sqrt{2N\log 2}\sum_{k=0}^K 2^{-kH}.
  \label{eq:B-Vmin}
\end{equation}
(This is exact for the GREM.) For $H>0$ the sum converges and only the top $O(1/H)$ scales contribute, $V_{\min}\sim-\sigma r_0^H\sqrt N$ with no $\log r_0$; for $H\to0$ all scales contribute equally and one recovers the log-correlated freezing result $V_{\min}\sim-\sigma\sqrt N\log r_0$~\cite{fyodorov2008freezing}.

Now, consider two low-lying minima at distance $\sim r_k$. Their common coarse-scale contributions $V_j$, $j<k$, cancel from their energy difference. The competition at scale $r_k$ is thus among the $2^N$ sub-cells of scale $r_k$ of a common parent cell of scale $r_{k-1}$, i.e. among
\begin{equation}
  \mathcal M = 2^N = \e^{cN}
  \label{eq:B-M}
\end{equation}
siblings ($c=\log 2$), whose level-$k$ values are i.i.d.\ Gaussians of variance $a_k^2$. (For $k=0$ the parent is the whole domain; on the sphere the $2^N$ orthants provide a natural top-level partition into cells of size $\sim L\sqrt N$.) Standard extreme-value statistics (EVT) \cite{bouchaud1997universality} for $\mathcal M$ i.i.d.\ Gaussians of variance $a_k^2$ predicts that the minimum sits at $-a_k\sqrt{2\log\mathcal M}$, while the second minimum sits at approximately the same location with spacing
\begin{equation}
  \Egap(r_k) \;\sim\; \frac{a_k}{\sqrt{2\log\mathcal M}} \;=\; \frac{\sigma r_k^H}{\sqrt{2N\log 2}}.
  \label{eq:B-EVT-gap}
\end{equation}
The relative gap between the top two is $1/(2\log\mathcal M) = 1/(2N\log 2)$, small in $N$: the two best cells are typically extremely close in energy compared to their absolute depth $\sim\sigma r_k^H\sqrt N$.

Fine-scale contributions $V_j$ ($j> k$) add, to each sibling, the (independent) minimum over its own sub-tree. These shifts have fluctuations $O(\sigma r_{k+1}^H/\sqrt N)$, at the same order as the gap, meaning that they don't change the scaling (nor the exponential intensity of the extreme-value process, see the freezing argument below). The result is Eq.~\eqref{eq:B-gap}, up to an $O(1)$ prefactor. The prefactor $1/\sqrt{2c}$ absorbs the $H$-dependent constants and the log-density $c$ of candidates per scale.

Applying the canonical scaling $\sigma = \tsigma N^{(1-H)/2}$, $r_0 = L\sqrt N$ with $\tsigma, L = O(1)$ [Eq.~\eqref{eq:canonical}], the natural energy unit is
\begin{equation}
  \sigma r_0^H = \tsigma\,L^H\,\sqrt N,
  \label{eq:B-unit}
\end{equation}
so the ground-state energy is extensive, $V_{\min}/N=O(\tsigma L^H)$. The extreme-value spacing becomes $\Egap(r_0) \sim \sigma r_0^H/\sqrt N = \tsigma L^H = O(1)$, i.e., the gap to a distant secondary minimum is $O(1)$ independently of $N$ and the canonical scaling sits precisely at the ``generic degeneracy'' scaling. The near-degeneracy probability that the gap falls within an $O(1)$ window correspondingly saturates, $P(r_0,N) \sim \sqrt N/(\sigma r_0^H) = 1/(\tsigma L^H) = O(1)$, meaning distant $O(1)$-degenerate secondary minima are generic rather than rare. This is the regime in which the problem is genuinely glassy and in which the replica analysis of Sec.~\ref{sec:equilibrium} applies. Here, the probability $\propto 1/\tsigma$ is the counterpart of the top Parisi breakpoint $u_0\propto 1/\tsigma$ (probability that two configurations lie in different top-level branches), see Eq.~\eqref{eq:B-x0} below.

A single level is effectively a REM of $2^N$ Gaussians of variance $a^2$, and thus has near its minimum $-a\sqrt{2N\log 2}$ a density of levels growing as $\e^{f/T_c}$ with $T_c=a/\sqrt{2N\log 2}$. It freezes below $T_c$, where its Boltzmann weights become Poisson--Dirichlet with parameter $u =T/T_c$. In the GREM the variance per unit entropy $a_k^2/(N\log 2)$ decreases with depth, so each generation freezes separately \cite{derrida1986solution}, generation $k$ at
\begin{equation}
  T_k=\frac{a_k}{\sqrt{2N\log 2}},
  \label{eq:B-Tk}
\end{equation}
with $T_0>T_1>\dots$, meaning the coarsest level freezes first upon cooling. In the canonical scaling, $a_k = \tsigma L^H\sqrt N\,2^{-kH}$ and
\begin{equation}
  T_k=\frac{\tsigma}{\sqrt{2\log 2}}\left(\frac{r_k}{\sqrt N}\right)^H,
  \label{eq:B-Tk-canonical}
\end{equation}
independent of $N$. In our $T=1$ setting, generation $k$ is frozen iff $T_k>1$, i.e.
\begin{equation}
  r_k \;>\; \ell_1^{\rm GREM}:=\sqrt N\,(2\log 2)^{1/(2H)}\,\tsigma^{-1/H},
  \label{eq:B-ell1}
\end{equation}
which reproduces the scaling $\ell_1=\sqrt{N}D_1\propto\sqrt N\tsigma^{-1/H}$ of the short-distance plateau [Eq.~\eqref{eq:q1}]. The top level is frozen iff $\tsigma L^H>\sqrt{2\log 2}$, the GREM counterpart of the strong-disorder condition $\tsigma>\tsigma_\AT\propto L^{-H}$, otherwise no level is frozen and the measure is replica symmetric.

The Parisi weight of a frozen generation is $u^{(k)}=T/T_k$; at $T=1$ and with $D=r_k/\sqrt N$,
\begin{equation}
  u(D)=\frac{\sqrt{2\log 2}}{\tsigma}\,D^{-H},
  \label{eq:B-xk}
\end{equation}
the same scaling $D^{-H}/\tsigma$ as Eq.~\eqref{eq:xq}. In the cascade, $u^{(k)}$ is the probability that two configurations drawn from the Gibbs measure lie in different generation-$k$ cells. Therefore, the last common ancestor has $\Pr[k^\star_{\alpha\gamma}=k]=u^{(k+1)}-u^{(k)}\propto2^{kH}$, and, exactly as in App.~\ref{app:A-pair}, $\rho(\ell)\propto\ell^{-1-H}$ on $\ell_1^{\rm GREM}<\ell<L\sqrt N$. The weight of the coarsest generation,
\begin{equation}
  u^{(0)}=\frac{\sqrt{2\log 2}}{\tsigma L^H}\;\propto\;\frac1\tsigma ,
  \label{eq:B-x0}
\end{equation}
is the counterpart of the lower breakpoint $u_0\propto 1/(\tsigma D_0^{H})$ of Eq.~\eqref{eq:Du}, meaning that at strong disorder the measure is condensed already at the coarsest level.

We can finally make the sibling gap $\Egap \sim T/u$ explicit. Within a frozen generation $k$, the free energies of the $2^N$ sibling cells (each the level-$k$ increment plus the free energy of its own sub-tree), measured from the lowest, form near the bottom a Poisson process of intensity $\propto\e^{f/T_k}=\e^{u^{(k)}f/T}$. There, the independent sub-tree shifts do not change the exponent of the intensity, because the deeper levels freeze at lower temperatures ($T_{k+1}<T_k$, so the required exponential moments are finite). This is Eq.~\eqref{eq:A-PPP} with an explicit $u$, and the typical sibling gap is
\begin{equation}
  \Egap(r_k)\sim\frac{T}{u^{(k)}}=T_k=\frac{\sigma r_k^H}{\sqrt{2N\log 2}},
  \label{eq:B-sibling-gap}
\end{equation}
reproducing both the EVT estimate \eqref{eq:B-EVT-gap} and Eq.~\eqref{eq:gap}.

Note that this real-space GREM captures exponents and scales but not prefactors, and it does not reproduce the atom of weight $H/2$ at $q_1$. Below $\ell_1^{\rm GREM}$, the levels are liquid, and with a continuum of levels $u$ reaches $1$ continuously instead of saturating at $u_1=(2-H)/2$. Overall, this dyadic construction mirrors Sinai's multifractal decomposition of Burgers flow with self-similar initial data~\cite{sinai1992statistics} and is similar to the $\beta$-model of turbulence~\cite{frisch1978simple,benzi1984multifractal}.

\section{Zero temperature curvature of the foam}
\label{app:foam-mobius}

In the Hopf--Lax limit, $W_\infty(\by)=\min_{\bx}U(\bx;\by)$, the Hessian of $W_\infty$ can be related directly to the Hessian of $V$ at the minimiser. Let $\bx^\star(\by)$ denote the minimiser of $U(\bx;\by) = V(\bx) + \kappa|\by-\bx|^2/2$ at fixed $\by$, assumed unique; the stationarity condition reads $\nabla V(\bx^\star) = \kappa(\by-\bx^\star)$.
Differentiating implicitly with respect to $\by$,
\begin{equation}
\bigl[\nabla\nabla V(\bx^\star) + \kappa \mathbf I\bigr] \frac{\partial\bx^\star}{\partial\by} = \kappa \mathbf I,
\end{equation}
and using $\nabla W_\infty = \kappa(\by-\bx^\star)$, so that $\nabla\nabla W_\infty = \kappa\mathbf I - \kappa\,\partial\bx^\star/\partial\by$, one obtains
\begin{equation}
\nabla\nabla W_\infty(\by) = \kappa \mathbf I - \kappa^2\bigl[\nabla\nabla V(\bx^\star) + \kappa\mathbf I\bigr]^{-1},
\label{eq:mobius}
\end{equation}
which is nothing but the zero-temperature form of Eq.~\eqref{eq:hessian}.

In the eigenbasis of $\nabla\nabla V(\bx^\star)$, with eigenvalues $\omega_i$, the curvatures of $W_\infty$ are given by the map
\begin{equation}
\Lambda_i=\frac{\kappa\omega_i}{\omega_i+\kappa}.
\end{equation}
Minima of $V$ are thus mapped onto curvatures $0<\Lambda_i<\kappa$, so that $W_\infty$ is always softer than the bare harmonic cost, and soft directions of $V$ remain soft directions of $W_\infty$. Conversely, $W_\infty$ becomes concave in a given direction only where the minimiser sits in a region of negative curvature of $V$, $-\kappa<\omega_i<0$. The map diverges at $\omega_i=-\kappa$, where the Hessian of $V(\bx)+\kappa|\by-\bx|^2/2$ at the minimiser acquires a zero mode. There, $\bx^\star(\by)$ can no longer be followed smoothly as $\by$ moves, and generically the minimum merges with a neighbouring saddle and disappears---i.e. undergoes a saddle-node bifurcation. The walls of $W_\infty$ are not located at these points, however. As $\by$ moves, the minimiser $\bx^\star(\by)$ follows the lowest of the local minima of $V(\bx)+\kappa|\by-\bx|^2/2$, and a wall is crossed when another local minimum becomes lower. There $\bx^\star$ jumps from one minimum to the other while both still exist, so that $W_\infty$, the lowest value, stays continuous while its gradient jumps. The divergence of the map instead signals the disappearance of a local minimum by a saddle-node bifurcation, which only occurs once this minimum has ceased to be the lowest one, i.e. beyond the wall, on a metastable branch. The two coincide only where a wall is born, at a point where the lowest minimum itself flattens and splits into two. This is the zero-temperature counterpart of the distinction between spinodal and binodal made in Sec.~\ref{sec:speciation}.

At $\beta=1$, this picture must be slightly amended. Within a pure state, treated as Gaussian, the Hessian of $V$ is to be replaced by the stiffness $\mathbf K^\alpha=(\mathbf C^\alpha)^{-1}$ of the state introduced in Sec.~\ref{sec:speciation}. As it is positive definite, the curvatures $\Lambda_i=\kappa\lambda_i/(\lambda_i+\kappa)$ are positive and the map does not diverge. The many shocks generated by the roughness of $V$ below the width $\ell_1$ of the states lie inside them, and are washed out by the thermal fluctuations. At $\beta=1$, the walls therefore arise only from the competition between pure states, which is precisely what the smoothed minimum~\eqref{eq:cells} describes.

\section{State attribution in the speciation regime}\label{app:attribution}

\subsection{Rigid-state posterior}\label{app:attribution-posterior}\label{app:attribution-check}

Drawing $\bx_0\sim P_0$ amounts to choosing a pure state $\alpha$ with probability $w_\alpha\propto\e^{-F_\alpha}$ and sampling $\bx_0=\bm m^\alpha+\bm\zeta_\alpha$ within it. As in Sec.~\ref{sec:attribution-speciation}, we treat the states as rigid, neglecting the intra-state fluctuations $\bm\zeta_\alpha$, which are sub-dominant with respect to the inter-state ones (Sec.~\ref{sec:covariance}) and are only resolved at $\kappa=O(1)$ (Sec.~\ref{sec:short}). For a noised point $\by=\bm m^\alpha+\bz/\sqrt\kappa$ planted in $\alpha$, with $\bz\sim\mathcal N(\bm0,\mathbf I_N)$, the posterior weight of a state $\gamma$ is
\begin{equation}
\pi_\gamma(\by,\kappa)
=\frac{w_\gamma\langle\e^{\kappa\by\cdot\bx}\rangle_\gamma}
{\sum_{\gamma'} w_{\gamma'}\langle\e^{\kappa\by\cdot\bx}\rangle_{\gamma'}} \simeq 
\frac{w_\gamma\,\e^{\tkappa Q_{\alpha\gamma}+\sqrt{\tkappa}\,g_\gamma}}{\sum_{\gamma'} w_{\gamma'}\,\e^{\tkappa Q_{\alpha\gamma'}+\sqrt{\tkappa}\,g_{\gamma'}}},
\label{eq:attribution-posterior}
\end{equation}
with Gaussian fields $g_\gamma=\bz\cdot\bm m^\gamma/\sqrt N$, $\mathbb E\,g_\gamma g_{\gamma'}=Q_{\gamma\gamma'}$. The first equality is exact on the sphere, where $|\bx|^2=NL^2$. The second replaces each state by its centre and uses $\kappa\,\bm m^\alpha\cdot\bm m^\gamma=\tkappa Q_{\alpha\gamma}$, the terms $\kappa|\bm m^\gamma|^2/2=\tkappa q_1/2$ being the same for all states. The posterior thus depends on $N$ only through $\tkappa$. Taking the ratio of the weights of a competitor $\gamma$ and of the planted state $\alpha$, with $Q_{\alpha\alpha}=q_1$,
\begin{equation}
\log\frac{\pi_\gamma}{\pi_\alpha}=-(F_\gamma-F_\alpha)-\tkappa\,(q_1-Q_{\alpha\gamma})+\sqrt{\tkappa}\,(g_\gamma-g_\alpha) .
\end{equation}
With $D^2_{\alpha\gamma}=2(q_1-Q_{\alpha\gamma})$, the second term is the penalty $\tkappa D^2_{\alpha\gamma}/2$, while $g_\gamma-g_\alpha$ is Gaussian with zero mean and variance $Q_{\gamma\gamma}+Q_{\alpha\alpha}-2Q_{\alpha\gamma}=D^2_{\alpha\gamma}$, i.e. $g_\gamma-g_\alpha=D_{\alpha\gamma}\,\xi$ with $\xi\sim\mathcal N(0,1)$. This gives the log-odds ratio~\eqref{eq:log-odds} of the main text.

\subsection{Denoising along independent splits}\label{app:splits}

We detail here the independent-split approximation leading to Eq.~\eqref{eq:independent-splits}. Consider a single split between the branch of the planted state $\alpha$ and a branch $\gamma$ whose centres have overlap $q$ with it, with the two states at squared distance $D^2_{\alpha\gamma}=2(q_1-q)$. By Eq.~\eqref{eq:log-odds}, the observation shifts the free-energy difference $f= F_\gamma-F_\alpha$ between the two branches by $\delta f=\tkappa D^2_{\alpha\gamma}/2-\sqrt{\tkappa} D_{\alpha\gamma}\,\xi$ in favour of the branch which contains the planted state. Given a split with free-energy difference $f$, the probability of a wrong guess is the prior weight $1/(1+\e^{-f})$ of the branch of the planted state times the posterior weight $1/(1+\e^{f+\delta f})$ that the other branch keeps after the noisy observation. Placing ourselves among the balanced splits, $f$ is spread uniformly, and integrating this product over $f$ gives $\delta f
/(\e^{\delta f}-1)$. Writing $c=\tkappa D^2_{\alpha\gamma}/2=\tkappa(q_1-q)$ and averaging over the noise, the probability of a wrong guess at the level $q$ is thus reduced by a factor
\begin{equation}
  G(c)=\mathbb E_\xi\left[\frac{c-\sqrt{2c}\,\xi}{\e^{c-\sqrt{2c}\,\xi}-1}\right],
  \label{eq:F-two-state}
\end{equation}
compared with a guess drawn from the prior, for which $\delta f=0$. It decreases from $G(0)=1$ to $G(c)\propto c^{-1/2}\e^{-c/4}$ at large $c$.

It remains to count the splits along the branch of the planted state. By definition of the Parisi function, a guess drawn from the prior has an overlap smaller than $q$ with the true state with probability $u(q)$, so it leaves the branch of the true state at the level $q$ with probability $P(q)\,\dd q$, with $P(q)=\dd u/\dd q$ the distribution of overlaps of Sec.~\ref{sec:covariance}, including its peak of weight $u_0$ at $q_0$. Such a guess costs $q_1-q$ in the second form of Eq.~\eqref{eq:centre-mmse}, $\Delta_c=\mathbb E\sum_\gamma\pi_\gamma(q_1-Q_{\alpha\gamma})$, and treating each level as an independent split gives Eq.~\eqref{eq:independent-splits}.

\subsection{Monte Carlo sampling of the posterior over states}\label{app:mc}

The Monte Carlo points of Fig.~\ref{fig:two-scale} evaluate the centre error~\eqref{eq:centre-mmse} for rigid states, without the independent-split approximation leading to Eq.~\eqref{eq:independent-splits}. Indeed, the posterior~\eqref{eq:attribution-posterior} only involves the weights of the states and their mutual overlaps, so that $N$ never enters. Here, a sample therefore consists of a tree of states with its weights, a planted state drawn with its weight, and one realisation of the noisy Gaussian fields $g_\gamma$. We describe these steps in turn, and then the truncation that makes the tree finite.

\begin{enumerate}[label=(\roman*)]
  \item \emph{Discretised tree.} As in a $K$-step replica-symmetry-breaking solution, the continuous Parisi function is replaced by a staircase with $K$ levels, $u(q)=u^{(k)}$ for $q^{(k-1)}\le q<q^{(k)}$, with $q^{(0)}=q_0$, $q^{(K)}=q_1$, $u^{(1)}=u_0$ and $u^{(K)}=u_1$, so that the atoms $u_0$ at $q_0$ and $1-u_1$ at $q_1$ are kept. The $u^{(k)}$ are the values of $u$ at $K$ points log-spaced in $D$ from $D_0$ to $D_1$, and the overlaps $q^{(k)}$, $0<k<K$, sit at the geometric midpoints in $D$. $K$ is chosen such that $u$ changes little from one level to the next, $K=10$ for Fig.~\ref{fig:two-scale}. The tree then has $K$ levels below its root, and two states whose branches separate at depth $k$ have overlap $Q_{\alpha\gamma}=q^{(k)}$.
  \item \emph{Weights.} In the picture of Sec.~\ref{sec:potential}, the children of a node at depth $k-1$ are sub-clusters whose free energies are distributed like the lowest levels of a random energy model, with a density growing as $\e^{u^{(k)}F}$, the weight of each child being its Boltzmann weight normalised within its parent. Rather than drawing these free energies, we generate the weights of the children one after the other, by breaking off random fractions of the mass of the parent. The first child receives a fraction $V_1$ of this mass, the second a fraction $V_2$ of what remains, and so on, with independent $V_j\sim\mathrm{Beta}(1-u^{(k)},\,ju^{(k)}-u^{(k-1)})$ and $u^{(0)}=0$. This stick-breaking construction reproduces exactly the law of the weights of the Parisi tree~\cite{pitman1997two,bolthausen1998ruelle}, the children appearing in size-biased order, i.e. each new child is drawn among the remaining ones with a probability proportional to its weight. For $u^{(k)}\ll1$, $V_1$ is close to one, so that a single child carries almost all the mass of its parent, which is the condensation discussed in the main text. The leaves at depth $K$ are the pure states, and their masses are the weights $w_\gamma$. (Note that this is different from the real-space GREM construction described above that goes all the way down to the UV cutoff.)
  \item \emph{Planted state.} The planted state $\alpha$ must be drawn with probability $w_\alpha$. Equivalently, one can descend the tree from its root, choosing at each node the next branch with a probability proportional to its weight. This is exactly how the first child of the stick-breaking is chosen, so that the planted state is the leaf reached by always following the first child, its competitors at each level being the other children of the nodes along this path. Building the tree around the planted path allows us to keep this path exactly, however small its weight, while truncating its light competitors.
  \item \emph{Fields and posterior.} Each edge from depth $k-1$ to depth $k$ carries an independent Gaussian increment of variance $q^{(k)}-q^{(k-1)}$, and $g_\gamma$ is the sum of the increments along the path of $\gamma$, so that $\mathbb E\,g_\gamma g_{\gamma'}=Q_{\gamma\gamma'}-q_0$. The missing part $q_0$ is common to all states and cancels in $\pi_\gamma$. The posterior then follows from Eq.~\eqref{eq:attribution-posterior}, i.e. from the log-weights $\log w_\gamma+\tkappa Q_{\alpha\gamma}+\sqrt{\tkappa}\,g_\gamma$, and one draw of the fields serves for all values of $\tkappa$. The estimator $q_1-\sum_\gamma\pi_\gamma Q_{\alpha\gamma}$ has mean $\Delta_c$ by the second form of Eq.~\eqref{eq:centre-mmse}, and is averaged over trees and field draws. 
  \item \emph{Truncation.} Each node has infinitely many children, almost all of them extremely light, so the tree must be truncated. Along a branch that leaves the planted path at depth $j$, children lighter than a fraction $w_{\rm rel}$ of the mass of the planted path at depth $j$ are merged, together with the mass not yet broken off, into a single lump per parent, which keeps its exact total weight and its exact overlap $q^{(j)}$ with the planted state. The competitors of the planted state at the finest level, which carry the error at large $\tkappa$, are kept down to a much smaller fraction $w_{\rm leaf}$ of the mass of their parent. A lump attached to a node at depth $d\geq j$ carries the field of that node, while the increments below it are unresolved. They are replaced by the factor $\e^{\tkappa(q_1-q^{(d)})/2}$ that they contribute on average. The truncation error should vanish as $w_{\rm rel},w_{\rm leaf}\to0$. Monte Carlo points are shown where the result no longer changes when $w_{\rm leaf}$ is lowered by a factor 10, down to $10^{-7}$, while we fix $w_{\rm rel}=10^{-5}$.

\end{enumerate}

\section{Replica computations at fixed $\kappa$}\label{app:planted}

This appendix gives the replica computations behind Sec.~\ref{sec:short} for the internal measure $p(\bx\mid\by,\kappa)$ at fixed $\kappa>0$, after $N\to\infty$. We first treat the naive case, in which $\by$ is independent of the disorder, and then the planted case, where $\by$ is a noisy version of an equilibrium configuration and the unknown is the posterior width $\Delta(\kappa)=L^2-q_\post$, i.e. the per-coordinate MMSE.

\subsection{Naive independent pinning}\label{app:planted-independent}

A natural first attempt is to consider $Z(\by,\kappa)\propto\int\dd\bx\,\exp\big(-V(\bx)-\frac\kappa2|\by-\bx|^2\big)$ for a $\by$ fixed \emph{independently of the disorder}, and to compute $\overline{\log Z(\by,\kappa)}$ with replicas as for the equilibrium problem. Since the data lie on the sphere $|\bx|^2 = NL^2$, the internal measure is $\propto\exp[-V(\bx)+\kappa\by\cdot\bx]$, corresponding to the original problem in an external field directed along $\by$, of strength $h^2=|\kappa\by|^2/N=\kappa^2r^2$, with $r^2=|\by|^2/N$. By isotropy, the result depends only on $h$, and averaging over the direction of $\by$ turns the field into a channel term acting on the inner replicas only,
\begin{equation}
\mathcal S_h(Q)=\frac12\log\det Q+\frac12\sum_{a,b}\cB(D^2_{ab})+\frac{h^2}2\sum_{a,b}Q_{ab}.
\label{eq:F-indep-action}
\end{equation}

For the replica-symmetric ansatz $Q_{a\neq b}=q$, $\Delta=L^2-q$, the saddle-point equation becomes
\begin{equation}
\frac{q}{\Delta^2}
=
2^H\tsigma^2H\,
\Delta^{H-1}
+h^2.
\label{eq:F-RS-field}
\end{equation}
The field term being linear in $Q$, the replicon is unchanged, $\lambda_R=\Delta^{-2}-4\cB''(2\Delta)$, and the RS solution loses stability when
\begin{equation}
1
=
2^H\tsigma^2H(1-H)
\Delta^H,
\label{eq:F-indep-replicon}
\end{equation}
i.e. at $\Delta=\Done$ [Eq.~\eqref{eq:q1}]. Eliminating $q$ between Eqs.~\eqref{eq:F-RS-field} and \eqref{eq:F-indep-replicon} gives the AT field
\begin{equation}
h_{\rm AT}^2
=\frac{L^2}{\Done^2}-\frac{2-H}{1-H}\,\frac1\Done.
\label{eq:F-hAT}
\end{equation}
The internal measure is replica symmetric for $h^2>h_{\rm AT}^2$ and fRSB for $h^2<h_{\rm AT}^2$ ($h_{\rm AT}=0$ at $\tsigma=\tsigma_{\rm AT}$, where $\Done=L^2(1-H)/(2-H)$).

\begin{figure}[t]
    \centering
    \includegraphics[width=\linewidth]{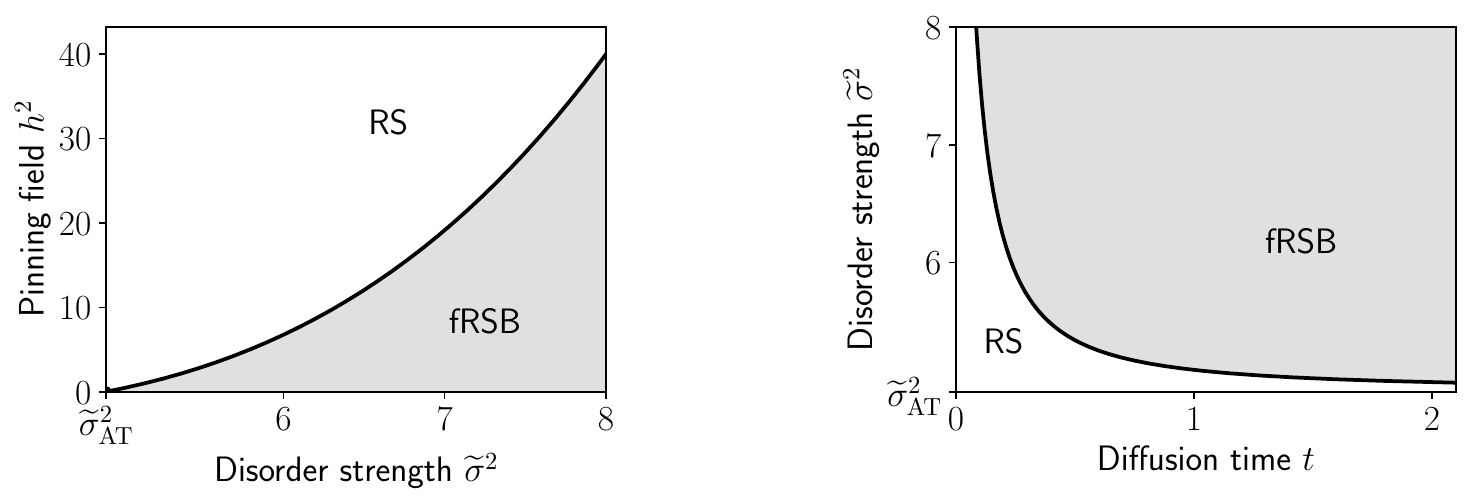}
    \caption{Independent (non-planted) pinning: replica-symmetry breaking of the internal measure $\propto\e^{-V(\bx)+\kappa\by\cdot\bx}$ for $\by$ independent of the disorder, at $H=1/2$ and $L=1$. Left: phase diagram in the plane of disorder strength $\tsigma^2$ and pinning field $h^2=|\kappa\by|^2/N$; the internal measure is RS above the AT line $h^2=h^2_{\rm AT}(\tsigma)$ of Eq.~\eqref{eq:F-hAT} and fRSB below it. Right: the same transition along a typical diffusion trajectory, $h^2=h^2_{\rm typ}(\kappa)=\kappa^2L^2+\kappa$ with $\kappa=1/(2t)$, in the plane of diffusion time $t$ and $\tsigma^2$; the region $t>t_{\rm AT}^{\rm int}(\tsigma)$ [Eq.~\eqref{eq:F-kappaAT-int}] would be fRSB. Only the strong-disorder regime $\tsigma^2\geq\tsigma^2_{\rm AT}$ is shown. This is \emph{not} the conditioning ensemble relevant for diffusion: for a planted $\by$, the internal measure is RS for every fixed $\kappa>0$ (App.~\ref{app:planted-stability}).}
    \label{fig:independent-pinning}
\end{figure}

Along a typical noisy trajectory, we have $\by=\bx_0+\bz/\sqrt\kappa$ which by concentration sits on the sphere $r^2\to L^2+1/\kappa$, hence
\begin{equation}
h_{\rm typ}^2(\kappa)=\kappa^2L^2+\kappa .
\label{eq:F-htyp}
\end{equation}
Substituting it into the AT criterion, $h_{\rm typ}^2(\kappa_{\rm AT}^{\rm int})=h_{\rm AT}^2$, gives
\begin{equation}
\kappa_{\rm AT}^{\rm int}
=
\frac{
\sqrt{1+4L^2h_{\rm AT}^2}-1
}{
2L^2
},
\label{eq:F-kappaAT-int}
\end{equation}
i.e. $t_{\rm AT}^{\rm int}=1/(2\kappa_{\rm AT}^{\rm int})$ and the phase diagram of Fig.~\ref{fig:independent-pinning}. At strong disorder $\kappa_{\rm AT}^{\rm int}\simeq1/\Done\propto\tsigma^{2/H}$, which is the disorder dependence of the end of the cascade, $\tkappa_c$, but at $\kappa=O(1)$ instead of $O(1/N)$. Clearly this contradicts the main result of our paper which is that speciation occurs entirely on the $\tkappa$ scale.

\subsection{The planted object and its action}\label{app:planted-object}\label{app:planted-action}

In order to obtain a result consistent with the Larkin picture, we must instead consider the strong correlation between the noisy observation and the disorder in the $\kappa = O(1)$ regime.

We set out to compute $\mathcal F(\kappa)$, Eq.~\eqref{eq:planted-object}, with $Z(\by,\kappa)=\int\dd\bx\,\e^{-V(\bx)+\kappa\by\cdot\bx}$. Constants $\kappa|\by|^2/2$ and $\kappa|\bx|^2/2$ have been dropped ($|\bx|^2=NL^2$), so that $Z$ defines the same internal measure as Eq.~\eqref{eq:internal}. In this planted computation, two replica tricks are needed, first to compute the normalisation $\mathcal Z^{-1}=\lim_{m\to0}\mathcal Z^{m-1}$, and then to do the quenched average $\log Z=\lim_{n\to0}(Z^n-1)/n$ for the inner free energy. Both are taken on
\begin{equation}
\Xi(m,n)=\overline{\mathcal Z^{m-1}\int\dd\bx_0\,\e^{-V(\bx_0)}\,\mathbb E_{\bz}\,Z(\by,\kappa)^n},
\end{equation}
which satisfies $\Xi(m,0)=\overline{\mathcal Z^m}\to1$, so that
\begin{equation}
\frac{\mathcal F(\kappa)}N=\lim_{n\to0}\lim_{m\to0}\frac1{nN}\big[\log\Xi(m,n)-\log\Xi(m,0)\big].
\label{eq:planted-limits}
\end{equation} Writing the powers as products of integrals introduces two sets of replicas: the \emph{plant} set $\mathsf P=\{\bx_0,\bx_1',\dots,\bx'_{m-1}\}$, $m\to0$ replicas of the equilibrium measure, and the \emph{inner} set $\mathsf I=\{\bx_1,\dots,\bx_n\}$, $n\to0$ replicas of the internal measure. Explicitly,
\begin{equation}
\Xi(m,n)=\overline{\int\prod_{c=1}^{m-1}\dd\bx'_c\,\e^{-V(\bx_c')}\int\dd\bx_0\,\e^{-V(\bx_0)}\,\mathbb E_{\bz}\prod_{a=1}^n\int\dd\bx_a\,\e^{-V(\bx_a)+\kappa\,\bx_a \cdot(\bx_0+\bz/\sqrt{\kappa})}}\,,
\label{eq:F-Xi}
\end{equation}

We start by integrating over $\bz$,
\begin{equation}
\mathbb E_{\bz}\prod_{a\in \mathsf I}\e^{\kappa\,\bx_a \cdot(\bx_0+\bz/\sqrt{\kappa})}
=\exp\left[\kappa\sum_{a\in \mathsf I}\bx_0\cdot\bx_a+\frac\kappa2\left|\sum_{a\in \mathsf I}\bx_a\right|^2\right]
=\exp\left[\frac\kappa2\left|\bx_0 + \sum_{a\in \mathsf I}\bx_a\right|^2-\frac{\kappa}{2}|\bx_0|^2\right].
\label{eq:F-channel}
\end{equation}
Since $|\bx_0|^2=NL^2$ is a constant, the channel couples the planted replica $\bx_0$ and the inner replicas \emph{symmetrically}, meaning that $\bx_0$ is a posterior sample, exchangeable with $\bx_1,\dots,\bx_n$. This is the replica-level expression of the Nishimori identity, and is what makes the computation tractable.

After the Gaussian disorder average over all $m+n$ replicas (as for the equilibrium problem, App.~\ref{app:replica}), $\Xi(m,n)\asymp\int_{\mathcal Q_{m+n}}\dd Q\,\e^{N\mathcal S(Q)}$ with
\begin{equation}
\mathcal S(Q)=\frac12\log\det Q+\frac12\sum_{i,j\in \mathsf P\cup \mathsf I}\cB(D^2_{ij})+\frac\kappa2\sum_{i,j\in\{0\}\cup \mathsf I}Q_{ij}-\frac\kappa2L^2,
\label{eq:F-action}
\end{equation}
where $Q$ is the $(m+n)\times(m+n)$ overlap matrix between \emph{all} the configurations, both from the planted and the inner set, $D_{ij}^2=2(L^2-Q_{ij})$, and $\mathcal Q_{m+n}$ is the set of positive-definite matrices with diagonal $L^2$. The indices $i,j$ therefore run over all replicas. In the following, indices $a,b$ will be reserved for inner replicas, and $c$ for the plant ones other than $\bx_0$. The first two terms are the equilibrium action for $m+n$ replicas, while the channel adds a term linear in $Q$ on the $n+1$ exchangeable replicas $\{0\}\cup\mathsf I$.

Finding ourselves with these two sets of replicas, we decompose the overlaps with the following mixed replica-symmetry ansatz:
\begin{itemize}
\item The plant block $Q^{\mathsf P}$ ($m\times m$): this block inherits all of the properties of the equilibrium measure described in Sec.~\ref{sec:potential}. As such, in the strong disorder regime it is natural to expect full replica-symmetry breaking, with diagonal overlaps $L^2$ and a continuous function $u(q)$ on $[q_0,q_1]$ as we take $m\to 0$.
\item The inner block $Q^{\mathsf I}$ ($n\times n$): following speciation, we expect it to be replica symmetric, with $Q_{ab}=q_\post$ for $a\ne b$, with $q_\post \geq q_1$ (we will equivalently write $\Delta := L^2 - q_\post \leq \Delta_1$).
\item The cross block: within the posterior RS ansatz, the Nishimori identity gives $Q_{a0}=q_\post$, equal to the overlap of two distinct inner replicas. We further assume that the inner replicas follow the branch containing $\bx_0$, inheriting its overlaps with the other plant replicas: $Q_{ac}=Q_{0c}$ for $c\in\mathsf P\setminus\{0\}$ and $a\in\mathsf I$. These cross overlaps are bounded by $q_1$.
\end{itemize}
Ordering the replicas as $(\mathsf P,\mathsf I)$ with $\bx_0$ first, and writing $\bm q$ for the column $(q_\post,Q_{01},\dots,Q_{0,m-1})^{\top}$,
\begin{equation}
Q=\begin{pmatrix}Q^{\mathsf P}&\bm q\,\bm 1_n^{\top}\\\bm 1_n\bm q^{\top}&\Delta\,\mathbf I_n+q_\post\mathbf J_n\end{pmatrix},
\label{eq:planted-block}
\end{equation}
with $\mathbf J_n$ an $n\times n$ matrix of all ones. At leading order in $n$, the plant block is evaluated on its equilibrium saddle. Within this ansatz, the only remaining unknown is thus $q_\post(\kappa)$, which we determine by stationarity of the action.

\subsection{Evaluation of the action}\label{app:planted-evaluation}

We evaluate the action~\eqref{eq:F-action} on the mixed ansatz~\eqref{eq:planted-block}. Given $q_\post>q_1\geq Q_{0c}$, its cross block is also what ultrametricity of the triplet $(\bx_0,\bx_a,\bx'_c)$ forces, and the $O(n)$ shift of the plant block away from its equilibrium saddle contributes only at $O(n^2)$, by stationarity.

\paragraph{Determinant term.}
By the Schur complement,
\begin{equation}
\det Q =\det Q^{\mathsf P}\cdot\det\big[\Delta\,\mathbf I_n+(q_\post-\Upsilon)\mathbf J_n\big],
\end{equation}
with $\Upsilon:=\bm q^{\top}(Q^{\mathsf P})^{-1}\bm q$. Since $\det[\Delta\mathbf I_n+b\,\mathbf J_n]=\Delta^{n-1}(\Delta+nb)$,
\begin{equation}
\frac12\log\det Q=\frac12\log\det Q^{\mathsf P}+\frac n2\left[\log\Delta+\frac{q_\post-\Upsilon}{\Delta}\right]+O(n^2).
\label{eq:F-logdet-schur}
\end{equation}
The vector $\bm q$ is the $0$-th column of $Q^{\mathsf P}$ with its diagonal entry $L^2$ replaced by $q_\post$, i.e. $\bm q=Q^{\mathsf P}\bm e_0-\Delta\,\bm e_0$. Hence $\Upsilon=L^2-2\Delta+\Delta^2[(Q^{\mathsf P})^{-1}]_{00}$, and thus
\begin{equation}
\frac{q_\post-\Upsilon}{\Delta}=1-\Delta\,\cP,
\label{eq:F-Upsilon}
\end{equation}
where $\cP:=\lim_{m\to0}\big[(Q^{\mathsf P})^{-1}\big]_{00}$ is the diagonal element of the inverse equilibrium Parisi matrix in the replica limit.

\paragraph{Diagonal element of the inverse Parisi matrix.}
To evaluate $\cP$, recall from App.~\ref{app:replica} the function $\chi(q)=\Done+\int_q^{q_1} \dd p \, u(p)$ [Eq.~\eqref{eq:A-chi}], constant for $q\le q_0$, and the equilibrium saddle relations $\chi(q)^{-2}=4\cB''(2(L^2-q))$ for $q_0<q<q_1$, $\chi(q_1)=\Done$ and $q_0=-\cB'(D_0^2)/(2\cB''(D_0^2))$ [Eqs.~\eqref{eq:A-chi-saddle}--\eqref{eq:A-q0}]. In this fRSB setting it is a standard result that
\begin{equation}
\lim_{m\to 0} \frac1m \log \det Q^{\mathsf P} = \log \Done + \int_{q_0}^{q_1}\frac{\dd q}{\chi(q)} + \frac{q_0}{\chi(q_0)}.
\label{eq:F-logdet-parisi}
\end{equation}
By Jacobi's formula, a uniform shift of the diagonal $L^2$ at fixed off-diagonal entries gives $\partial_{L^2}\log\det Q^{\mathsf P}=\Tr(Q^{\mathsf P})^{-1}=m\,\cP$. Since $L^2$ enters \eqref{eq:F-logdet-parisi} only through $\Done$ and $\chi$, with $\partial_{L^2}\chi=1$,
\begin{equation}
\cP=\frac1{\Done}-\int_{q_0}^{q_1}\frac{\dd q}{\chi(q)^2}-\frac{q_0}{\chi(q_0)^2}.
\label{eq:F-inverse-parisi}
\end{equation}
On the equilibrium solution, $\chi^{-2}=4\cB''(D^2)$ on the continuum (with $\dd q=-\dd D^2/2$) and, by continuity of $\chi$, $q_0/\chi(q_0)^2=-2\cB'(D_0^2)$, so that the $\cB'(D_0^2)$ contributions cancel:
\begin{equation}
\int_{q_0}^{q_1}\frac{\dd q}{\chi(q)^2}+\frac{q_0}{\chi(q_0)^2}=2\cB'(D^2_0)-2\cB'(D^2_1)-2\cB'(D^2_0)=-2\cB'(D^2_1),
\end{equation}
finally leading to
\begin{equation}
\cP=\frac1{\Done}+2\cB'(2\Done).
\label{eq:F-P}
\end{equation}
The lower plateau and the entire continuum drop out, and the plant block enters the inner-block equations only through $\Done$. In terms of the stiffness $\mu$ of Sec.~\ref{sec:planted}, $\cP=\mu(\Done)$. Combining with Eqs.~\eqref{eq:F-logdet-schur}--\eqref{eq:F-Upsilon}, the determinant term contributes $\frac n2[\log\Delta+1-\Delta\,\mu(\Done)]+O(n^2)$, the $n$-independent $\frac12\log\det Q^{\mathsf P}$ cancelling in Eq.~\eqref{eq:planted-limits}.

\paragraph{Disorder and channel terms.}
We split the disorder sum as $\sum_{\mathsf P,\mathsf P}+2\sum_{\mathsf P,\mathsf I}+\sum_{\mathsf I,\mathsf I}$. The first is $n$-independent, so it again drops from $\log \Xi(m,n) - \log \Xi(m,0)$. The last is $n\cB(0)+n(n-1)\cB(2\Delta)$ by replica symmetry. In the cross sum, each inner replica contributes $\cB(2\Delta)$ from $\bx_0$ and $-\int_0^1\dd u\,\cB(D^2(u))$ from the other plant replicas, using the standard $m\to0$ rule $\sum_{c\ne0}f(Q_{0c})\to-\int_0^1 \dd u \, f(q(u))$. Altogether,
\begin{equation}
\frac12\sum_{i,j}\cB(D^2_{ij})=\text{const}+\frac n2\left[\cB(0)+\cB(2\Delta)-2\int_0^1\dd u\,\cB(D^2(u))\right]+O(n^2).
\label{eq:F-disorder-term}
\end{equation}
Here $D(u)$ includes the plateaus of the Parisi function, $D(u)=D_0$ for $u<u_0$ and $D(u)=D_1$ for $u>u_1$, Eq.~\eqref{eq:Du} giving it in between.
The channel term involves the $n+1$ replicas $\{0\}\cup\mathsf I$, all at mutual overlap $q_\post$:
\begin{equation}
\frac\kappa2\sum_{i,j\in\{0\}\cup \mathsf I}Q_{ij}-\frac\kappa2L^2=\frac\kappa2\big[(n+1)L^2+(n+1)n\,q_\post\big]-\frac\kappa2L^2=\frac{n\kappa}2(L^2+q_\post)+O(n^2).
\label{eq:F-channel-term}
\end{equation}

\paragraph{Resulting planted free entropy.}
Collecting the $O(n)$ coefficients, the planted log-partition function per site, $\mathcal F(\kappa)/N$, is the extremum over $\Delta$ of
\begin{equation}
\phi(\kappa,\Delta)=\frac12\log\Delta+\frac12-\frac{\Delta}{2}\left[\frac1{\Done}+2\cB'(2\Done)\right]+\frac12\cB(2\Delta)+\frac\kappa2\big(2L^2-\Delta\big)+C_{\rm eq},
\label{eq:F-phi}
\end{equation}
with $C_{\rm eq}=\tfrac12\cB(0)-\int_0^1\dd u\,\cB(D^2(u))$ an equilibrium constant. The structure is effectively that of a Franz--Parisi-type potential \cite{franz1995recipes} for the overlap with the reference configuration.

\subsection{Saddle point and stability}\label{app:planted-saddle}\label{app:planted-stability}

In the planted free entropy~\eqref{eq:F-phi}, the lower plateau and the whole continuum of the Parisi solution have dropped out, and the plant enters only through the width $\Done$ of its state. Its extremum over the single remaining direction $\Delta$ is conveniently written in terms of the within-state ``stiffness''
\begin{equation}
\mu(\Delta)=\frac1\Delta+2\cB'(2\Delta),
\end{equation}
for which $\partial_\Delta\phi=\tfrac12\big[\mu(\Delta)-\mu(\Done)-\kappa\big]$. Stationarity thus gives the self-consistent equation~\eqref{eq:planted-saddle}, $\mu(\Delta)=\mu(\Done)+\kappa$, i.e. pinning raises the within-state stiffness by exactly $\kappa$. For our long-range family, $2\cB'(2\Delta)=-2^H\tsigma^2H\Delta^{H-1}$, and using Eq.~\eqref{eq:q1} to substitute $2^H\tsigma^2\Done^HH=1/(1-H)$, the saddle-point equation takes the form~\eqref{eq:planted-LR} in terms of the normalised width $\tilde\Delta=\Delta/\Done$. For $H=\tfrac12$ its right-hand side reduces to $(1/\sqrt{\tilde\Delta}-1)^2$, which gives the closed form
\begin{equation}
\Delta(\kappa)=\frac{\Done}{(1+\sqrt{\kappa\Done})^2}.
\label{eq:F-Hhalf}
\end{equation}
In general, the solution saturates at $\Done$ as $\kappa\to0$, as expected by construction of the ansatz, while as $\kappa\to\infty$ the term $1/\tilde\Delta$ dominates the right-hand side for every $H$, so that $\Delta(\kappa)\simeq1/\kappa$.

\begin{figure}[t]
    \centering
    \includegraphics[width=0.65\linewidth]{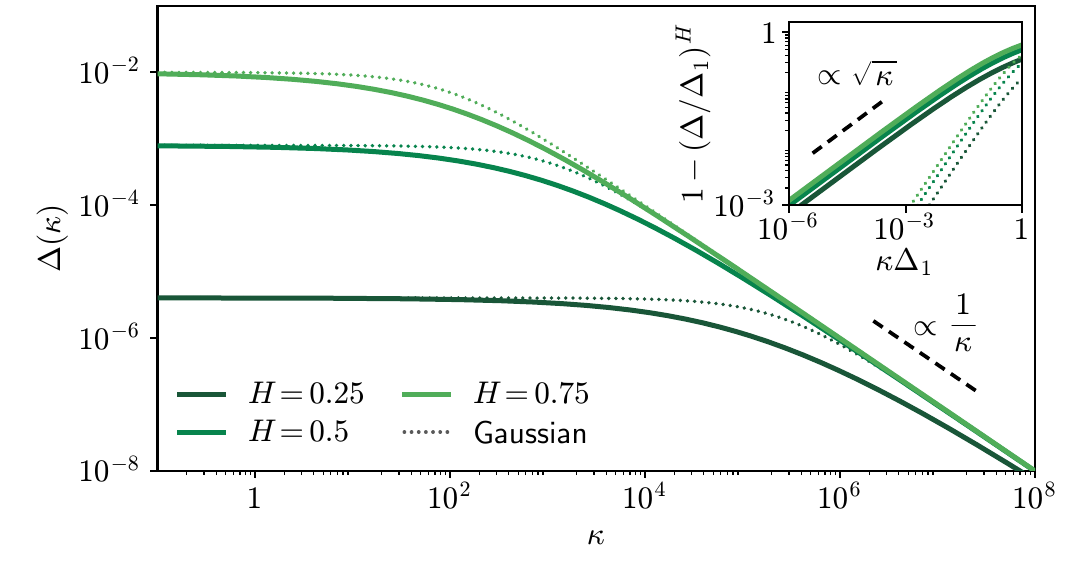}
    \caption{Posterior width, or equivalently denoising MMSE $\Delta(\kappa)$ on the planted replica-symmetric branch, Eq.~\eqref{eq:planted-saddle}, for $H=0.25,0.5,0.75$ at $\tsigma=10$. At small $\kappa$ it saturates at the thermal width $\Delta_1$ of the selected pure state, and at large $\kappa$, $\Delta\simeq1/\kappa$. Dotted, Gaussian state of the same width, $\Delta_1/(1+\kappa\Delta_1)$. Inset: $1-(\Delta/\Delta_1)^H$, equal to $\Delta^2\lambda_R^{\post}$ on the planted branch, versus $\kappa\Delta_1$. The planted branch leaves $\Delta_1$ as $\sqrt\kappa$, inherited from the marginality of the pure states (App.~\ref{app:planted-small-kappa}), whereas the Gaussian state leaves it linearly.}
    \label{fig:delta-kappa}
\end{figure}

By the Nishimori identity, $\Delta(\kappa)$ is the MMSE of the denoiser, shown in Fig.~\ref{fig:delta-kappa} for several values of $H$. For comparison, the dotted lines show a Gaussian state of the same width, for which the disorder term is absent, $\mu(\Delta)=1/\Delta$, so that Eq.~\eqref{eq:planted-saddle} gives $\Delta=\Done/(1+\kappa\Done)$, the MMSE of a Gaussian prior of variance $\Done$. Both start at $\Done$ and decay as $1/\kappa$, the true error lying below the Gaussian one at every $\kappa$ and converging to it at large $\kappa$.

It remains to check that the replica-symmetric inner block is stable, i.e. that the posterior at fixed $\by$ is indeed dominated by a single state for every $\kappa>0$. Here, the relevant fluctuations are those that break the symmetry between the inner replicas while leaving the rest of the ansatz untouched, namely symmetric perturbations $\delta Q$ of the inner block with $\delta Q_{aa}=0$ and $\sum_{b\in\mathsf I}\delta Q_{ab}=0$, the replicon fluctuations. Their vanishing row sums remove the rank-one coupling to the plant block, so that $Q^{-1}$ acts on them simply as $1/\Delta$, while the channel term, being linear in $Q$, does not contribute at second order. Expanding the determinant and disorder terms of Eq.~\eqref{eq:F-action} then gives $\mathcal S(Q+\delta Q)-\mathcal S(Q)=-\frac14\lambda_R^\post\Tr(\delta Q)^2$, with the replicon eigenvalue
\begin{equation}
\lambda_R^{\post}=\frac1{\Delta^2}-4\cB''(2\Delta)=-\mu'(\Delta),
\end{equation}
i.e. minus the slope of the within-state stiffness. Using Eq.~\eqref{eq:q1} to write $4\cB''(2\Delta)=\Done^{-H}\Delta^{H-2}$, its long-range form is Eq.~\eqref{eq:planted-replicon}. It is positive for every $0<\Delta<\Done$, and vanishes only at the equilibrium endpoint $\Delta=\Done$, where the pure state is marginal. The saddle point indeed remains in this interval. Writing $g(\Delta):=\mu(\Delta)-\mu(\Done)$, Eq.~\eqref{eq:planted-saddle} reads $g(\Delta)=\kappa$, with $g(\Done)=0$, $g(0^+)=+\infty$ and $g'=-\lambda_R^\post<0$ on $(0,\Done)$, so that there is a unique solution on this branch for every fixed $\kappa>0$, and it is stable. A root with $\Delta>\Done$, where $\lambda_R^\post<0$, would instead be unstable and would violate the assumption $q_\post\geq q_1$ of the ansatz. The scalar curvature is consistent with this, $\partial_{\Delta}^2\phi=-\lambda_R^{\post}/2<0$, the free entropy being maximised over the order parameter on the Nishimori line. Unlike the independently pinned problem of App.~\ref{app:planted-independent}, the planted saddle therefore has no AT instability at any finite $\kappa>0$, perfectly consistent with our global picture.

\subsection{\texorpdfstring{Small-$\kappa$ behaviour}{Small-kappa behaviour}}\label{app:planted-small-kappa}

The behaviour of the planted branch at small $\kappa$ is also consistent with the speciation picture of Sec.~\ref{sec:attribution-speciation}. As $\kappa\to0^+$ the planted branch tends to the width $\Done$ of a single pure state, while its stability margin closes, $\lambda_R^\post\to0$. This marginality is to be expected, since the selection of a single state, strict at any fixed $\kappa>0$, must give way to the competition between pure states of the regime $\kappa=\tkappa/N$. Here it takes the form of a $\sqrt\kappa$ departure from $\Done$. Expanding the saddle-point equation~\eqref{eq:planted-LR} around $\tilde\Delta=1$, $\kappa\Done=\frac H2(1-\tilde\Delta)^2+O\big((1-\tilde\Delta)^3\big)$, we indeed have
\begin{equation}
\Delta(\kappa)=\Done-\sqrt{\frac{2\Done^3}{H}}\,\sqrt\kappa+O(\kappa),\qquad \lambda_R^{\post}=\sqrt{\frac{2H}{\Done^3}}\,\sqrt\kappa+O(\kappa).
\label{eq:F-small-kappa}
\end{equation}
The $1/2$ exponent is due to the marginality of the pure states. It is thus a signature of the continuous replica-symmetry breaking brought by long-range correlations, $H$ entering only the prefactor. For short-range correlations, whose measure breaks replica symmetry in one step and whose states are not marginal, the onset would instead be linear, as for the Gaussian state (inset of Fig.~\ref{fig:delta-kappa}). This inheritance from marginality can also be seen through the so-called I-MMSE relations~\cite{guo2005mutual}. Their second-order form~\cite{payaro2009hessian} gives $-\partial_\kappa\Delta=N^{-1}\mathbb E\Tr\Chat^2$ for any prior, while differentiating Eq.~\eqref{eq:planted-saddle} with $\mu'=-\lambda_R^\post$ gives $\partial_\kappa\Delta=-1/\lambda_R^\post$. Hence $\lambda_R^\post=\big[N^{-1}\mathbb E\Tr\Chat^2\big]^{-1}$, and the diverging slope of the error as $\kappa\to0$ inherits the diverging spin-glass susceptibility of the soft modes of the pure state (Sec.~\ref{sec:covariance}).

\section{The source $\mathcal{J}$ of the perturbed reverse dynamics}\label{app:source}

In this appendix, we provide further details supporting Sec.~\ref{sec:source}.

\subsection{Exact representations and transport of $\mathcal{J}$}\label{app:G-exact}

\begin{enumerate}[label=(\alph*)]
  \item \emph{Posterior form}. From the Tweedie and Hessian identities \eqref{eq:tweedie}, \eqref{eq:hessian}, $|\nabla W|^2=\kappa^2|\by-\xhat|^2$ and $\nabla^2W=N\kappa-\kappa^2\Tr\Chat$, so that, using $|\by-\xhat|^2+\Tr\Chat=\langle|\by-\bx|^2\rangle_{\by,\kappa}$,
  \begin{equation}
    \mathcal{J}(\by,\kappa)=\kappa^2|\by-\xhat|^2+\kappa^2\Tr\Chat-N\kappa=\kappa^2\langle|\by-\bx|^2\rangle_{\by,\kappa}-N\kappa.
    \label{eq:G-S-posterior}
  \end{equation}
  The second form also follows directly from $\nabla^2_{\by}\e^{-\kappa|\by-\bx|^2/2}=(\kappa^2|\by-\bx|^2-N\kappa)\,\e^{-\kappa|\by-\bx|^2/2}$. It holds for any $P_0$.

  \item \emph{Conservation and fluctuations.} $\int P\,\mathcal{J}\,\dd\by=\int\nabla^2P\,\dd\by=0$. Equivalently, by the Nishimori identity, $\mathbb E|\by-\xhat|^2+\mathbb E\Tr\Chat=\mathbb E|\bz|^2/\kappa=N/\kappa$, i.e.
  \begin{equation}
    \mathcal{J}=\kappa^2\big(|\by-\xhat|^2-\mathbb E|\by-\xhat|^2\big)+\kappa^2\big(\Tr\Chat-\mathbb E\Tr\Chat\big).
    \label{eq:G-S-centred}
  \end{equation}
  Since $|\by-\xhat|^2\simeq|\bz|^2/\kappa$, the unconditional standard deviation of $\mathcal{J}$ is $\kappa\sqrt{2N}$, much larger than the mixture excess $\kappa^2\Tr\Chat_{\rm inter}$ (e.g.\ $\kappa\Gam/4$ for a wall at $\ell=\lL$, with $\Gam=O(1)$ in $N$). Walls and junctions are therefore visible in $\mathcal{J}$ only after conditioning on position as we do below. The quantity that detects them unconditionally is $\Tr\Chat=\kappa^{-1}(N-\nabla^2W/\kappa)$, i.e. $-\nabla^2W$, not $\mathcal{J}$ directly.
  \item \emph{Transport.} With $\partial_tW=\nabla^2W-|\nabla W|^2$ (the KPZ equation~\eqref{eq:KPZ} at $\beta=1$, forward time $t=1/(2\kappa)$) one has $\partial_tW=-\mathcal{J}$, and differentiating $\mathcal{J}=|\nabla W|^2-\nabla^2W$,
  \begin{equation}
    \partial_t\mathcal{J}+2\nabla W\cdot\nabla \mathcal{J}=\nabla^2\mathcal{J}.
    \label{eq:G-transport}
  \end{equation}
  $\mathcal{J}$ is thus a passive scalar, advected by the Burgers velocity $2\nabla W$ and diffused, with no source term.
\end{enumerate}

\subsection{Walls and junctions}\label{app:G-junctions}

To evaluate the source on the ``skeleton'' of the foam, we consider a point $\by$ where $K$ cells meet, i.e. a wall between two cells ($K=2$) or a junction of several. The centres of the $K$ competing states span an affine subspace of dimension $r=K-1$, and the $N-r$ remaining directions are orthogonal to all of them. In these transverse directions the posterior does not distinguish the competing states, $\by-\xhat$ is essentially the noise, $\kappa^2|\by-\xhat|_\perp^2\simeq(N-r)\kappa$, and the extensive terms of Eq.~\eqref{eq:G-S-posterior} cancel on average, as they do for $\mathcal{J}$ as a whole since $\int P \mathcal{J}=0$. In the $r$ spanned directions, $\by$ sits where the competing states have equal posterior weights, so that $\xhat$ is their barycentre, while $\Chat$ reduces to the inter-state covariance $\Chat_{\rm inter}$ of the competing centres, with eigenvalues $c_1,\dots,c_r$. The component $\Pi(\by-\xhat)$ of $\by-\xhat$ in this span vanishes only if the $K$ states also have equal prior weights, $\by$ being otherwise displaced from the barycentre by their free-energy differences. Averaging over the transverse noise,
\begin{equation}
  \mathbb E\big[\mathcal{J}\mid\text{junction}\big]=\kappa^2|\Pi(\by-\xhat)|^2+\kappa\sum_{i=1}^r(\kappa c_i-1)\;\geq\;-\sum_{i=1}^r\Lambda_i.
  \label{eq:G-S-junction}
\end{equation}
where $\Lambda_i=\kappa-\kappa^2c_i$ are the curvatures of $W$ along the span, Eq.~\eqref{eq:hessian}. On the skeleton, $\mathcal{J}$ is thus minus the total curvature of $W$ across the junction, plus a non-negative term that vanishes for equal prior weights. It is positive whenever the total curvature of $W$ along the span is negative. Concretely,
\begin{itemize}
  \item \emph{Wall} ($K=2$, $r=1$, $c=\ell^2/4$). For prior free energies differing by $\Delta F$, the wall is displaced from the bisector by $\Delta F/(\kappa\ell)$, so that $\mathbb E[\mathcal{J}\mid\text{wall}]=\kappa(\kappa\ell^2/4-1)+(\Delta F)^2/\ell^2=\kappa\big[\Gam/4-1+(\Delta F)^2/\Gam\big]$ at $\ell=\lL$. At the Larkin scale $\Delta F\sim\Egap\sim\Gam$, so that both terms are of the same order.
  \item \emph{Ultrametric (simplex) junction.} States branching at the same level of the ultrametric tree are mutually equidistant. $K$ of them at squared distance $\ell^2$ thus sit at the vertices of a regular simplex ($K\le N+1$), which spans a $(K-1)$-dimensional affine subspace, and the junction where their cells meet has codimension $K-1$ in $\by$-space. For equal weights,
  \begin{equation}
    \Chat_{\rm inter}=\frac1K\sum_{\alpha=1}^K(\bm m^\alpha-\bar{\bm m})(\bm m^\alpha-\bar{\bm m})^\top=\frac{\ell^2}{2K}\,\Pi_{K-1},
    \label{eq:G-simplex}
  \end{equation}
  with $\Pi_{K-1}$ the projector on their span. Up to the intra-state part, the curvature of $W$ is then $\kappa-\kappa^2\ell^2/(2K)$ in the $K-1$ in-simplex directions and $\kappa$ in the $N-K+1$ transverse ones, so that for $\kappa\ell^2>2K$ the junction is a maximum of $W$ in $K-1$ directions. The source reads
  \begin{equation}
    \mathbb E\big[\mathcal{J}\mid\text{simplex}\big]=\kappa(K-1)\left(\frac{\kappa\ell^2}{2K}-1\right),
  \end{equation}
  and is positive only for $K<\kappa\ell^2/2$. As $K$ grows, more directions become concave, but each of them less so, since the $K-1$ directions share the fixed spread of the centres, Eq.~\eqref{eq:G-trace}. The source is therefore largest for $K\simeq(\kappa\ell^2/2)^{1/2}$, where it reaches $\kappa[(\kappa\ell^2/2)^{1/2}-1]^2\simeq\kappa^2\ell^2/2$. At the cell scale, $\ell=\lL$, this is $K\simeq\sqrt{\Gam/2}$, with a source $\simeq\kappa\Gam/2$ of the same order as on a wall, $\kappa(\Gam/4-1)$, as stated in Sec.~\ref{sec:source}. The codimension of a junction thus controls the number of unstable directions of $W$, but not the sign of $\mathcal J$.
  \item \emph{General bound.} For any posterior over states,
  \begin{equation}
    \Tr\Chat_{\rm inter}=\frac12\sum_{\alpha,\gamma}\pi_\alpha\pi_\gamma|\bm m^\alpha-\bm m^\gamma|^2\;\le\;\frac12\,\max_{\alpha,\gamma}|\bm m^\alpha-\bm m^\gamma|^2,
    \label{eq:G-trace}
  \end{equation}
  half the squared diameter of the posterior support. By ultrametricity, this diameter does not grow with the number of competing states. A junction mixing several levels of the tree is a simplex of simplices, whose source is dominated by its coarsest active level. For states at a common scale $\ell\sim\lL$, the excess is thus bounded, $\kappa^2\Tr\Chat_{\rm inter}\le\kappa\Gam/2$, whatever the number of competing states and the codimension of the junction. In a sharp foam, junctions are therefore fed at most about twice as strongly as walls, and far more strongly than the cores of the cells. Being of order $\kappa\Gam$ rather than extensive in $N$, this excess is visible in $\mathcal J$ only once the radial fluctuations are averaged out at fixed position, as pointed out in App.~\ref{app:G-exact}.
\end{itemize}

\end{document}